\documentclass{krm}
\usepackage{amsmath}
\usepackage{amssymb}
\usepackage{bm}
  \usepackage{paralist}
  \usepackage{graphicx} 
 \usepackage[colorlinks=true]{hyperref}
\hypersetup{urlcolor=blue, citecolor=red}

\def\currentvolume{4}
 \def\currentissue{1}
  \def\currentyear{2011}
   \def\currentmonth{March}
    \def\ppages{361--384}
     \def\DOI{10.3934/krm.2011.4.361}

\theoremstyle{definition}

    \newcommand\beq{\begin{equation}}
    \newcommand\eeq{\end{equation}}
    \newcommand\beqa{\begin{eqnarray}}
    \newcommand\eeqa{\end{eqnarray}}
    \newcommand{\nn}{\nonumber\\}
    \newcommand{\gd}{\omega}
    \newcommand{\MM}{\mathcal{M}}
    \newcommand{\aone}{{(\text{I})}}
    \newcommand{\bone}{{(\text{II})}}
    \newcommand{\bonem}{{(\text{II},m)}}
    \newcommand{\bonezero}{{(\text{II},0)}}
    \newcommand{\boneone}{{(\text{II},1)}}
    \newcommand{\bonetwo}{{(\text{II},2)}}
    \newcommand{\Tot}{T_1^{(3)}}
    \newcommand{\poo}{p_1^{(1)}}
    \newcommand{\uot}{u_{x,1}^{(2)}}
    \newcommand{\qxoz}{q_{x,1}^{(0)}}
    \newcommand{\qxot}{q_{x,1}^{(2)}}
    \newcommand{\qyoz}{q_{y,1}^{(0)}}
    \newcommand{\qyot}{q_{y,1}^{(2)}}

\title[Couette--Poiseuille flow]
      {Non-Newtonian Couette--Poiseuille flow of\\ a dilute gas}

\author[Mohamed Tij and Andr\'es Santos]{}



 \email{mtij@fsmek.ac.ma}
 \email{andres@unex.es}

\begin{document}

\maketitle

\centerline{\scshape Mohamed Tij}
\medskip
{\footnotesize
 \centerline{D\'epartement de Physique, Universit\'e Moulay Isma\"il}
   \centerline{Mekn\`es, Morocco}
} 

\medskip

\centerline{\scshape Andr\'es Santos}
\medskip
{\footnotesize
 \centerline{Departamento de F\'{\i}sica, Universidad de Extremadura}
   \centerline{E-06071 Badajoz, Spain}
}



\begin{abstract}
The steady state of a dilute gas enclosed between two infinite parallel plates in relative motion and under the action of a uniform body force parallel to the plates is considered. The Bhatnagar--Gross--Krook model kinetic equation is analytically solved for this Couette--Poiseuille flow to first order in the force and for arbitrary values of the Knudsen number associated with the shear rate. This allows us to investigate the influence of the external force on the non-Newtonian properties of the Couette flow. Moreover, the Couette--Poiseuille flow is analyzed when the shear-rate Knudsen number and the scaled force are of the same order and terms up to second order are retained. In this way, the transition from the  bimodal temperature profile characteristic of the pure force-driven Poiseuille flow to the parabolic profile characteristic of the pure Couette flow through several intermediate stages in the Couette--Poiseuille flow are described. A critical comparison with the Navier--Stokes solution of the problem is carried out.
\end{abstract}

\section{Introduction}
\begin{figure}[tbp]
  \includegraphics[width=\columnwidth]{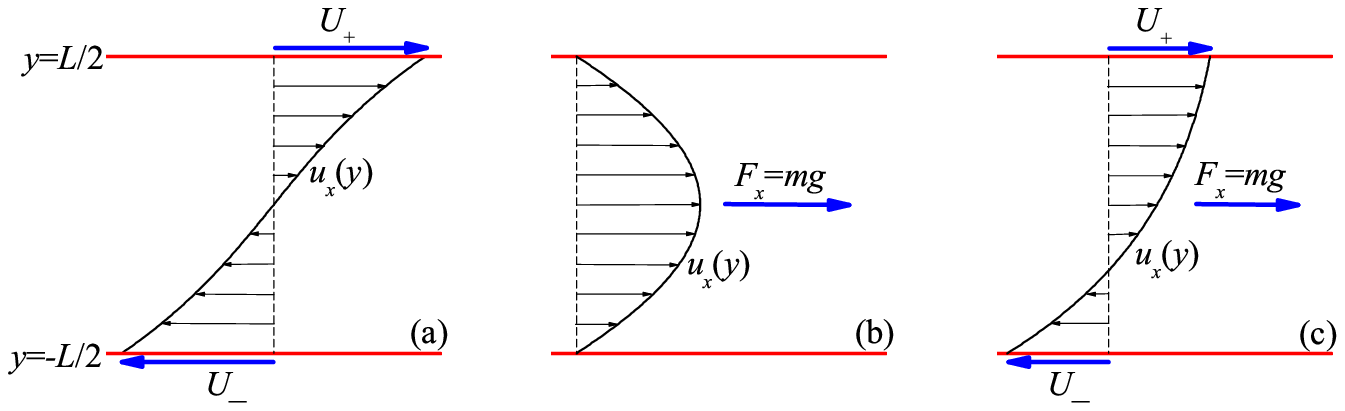}
\begin{center}
\caption{Sketch of (a) the Couette flow, (b) the force-driven Poiseuille flow, and (c) the Couette--Poiseuille flow.}\label{fig1}
  \end{center}
\end{figure}

Two paradigmatic stationary nonequilibrium flows are the plane Couette flow and the Poiseuille flow.
In the plane Couette flow the fluid (henceforth assumed to be a dilute gas) is enclosed between two infinite parallel plates in relative motion, as sketched in Fig.\ \ref{fig1}(a).
The walls can be kept at different or equal temperatures but, even if both wall temperatures are the same, viscous heating induces a temperature gradient in the steady state. If the Knudsen number associated with the shear rate is small enough the Navier--Stokes (NS) equations provide a satisfactory description of the Couette flow. On the other hand, as  shearing increases, non-Newtonian effects (shear thinning and viscometric properties) and deviations of Fourier's law (generalized thermal conductivity and streamwise heat flux component) become clearly apparent \cite{GS03}. These nonlinear effects have been derived from the Boltzmann equation for Maxwell molecules \cite{EF07,MN81,N83,S09,TS95}, from the Bhatnagar--Gross--Krook (BGK) kinetic model \cite{BSD87,GH97,SGB92}, and also from generalized hydrodynamic theories \cite{ST08,TTS09}. A good agreement with computer simulations \cite{GTRTS06,GTRTS07,KDSB89,MG98,MSG00,RC97} has been found. The plane Couette flow has also been analyzed in the context of granular gases \cite{TTMGSD01,VSG10}. In the case of plates at rest but kept at different temperatures, the Couette flow becomes the familiar plane Fourier flow, which also presents interesting properties by itself \cite{AMN79,GTRTS06,GTRTS07,KDSB89a,MASG94,N81,S09,SBG86,SBKD89}.

The Poiseuille flow, where  a gas is enclosed in a channel or slab and fluid motion is induced by a longitudinal pressure gradient, is a classical  problem in kinetic theory \cite{CS66,OSA89}.
Essentially the same type of flow field is generated when the pressure gradient is replaced by the action of a uniform
longitudinal body force $\mathbf{F} = mg\widehat{\mathbf{x}}$ (e.g., gravity), as illustrated in Fig.\ \ref{fig1}(b).  This force-driven
Poiseuille flow has received a lot of attention both from theoretical \cite{AS92,ATN02,ELM94,GVU08,HM99,M09,RC98,STS03,ST06,ST08,TTS09,TSS98,TS94,TS01,TS04,UG99,X03}
 and computational \cite{AC10,CA09,KMZ87,KMZ89,MBG97,RC98,TE97,TTE97,ZGA02} points of view.
This interest has been mainly motivated by the fact that the force-driven Poiseuille flow provides a nice example illustrating the limitations of the NS
description in the \emph{bulk} domain (i.e., far away from the boundary layers). In particular, while the NS equations predict a temperature profile with a flat maximum at the center, computer simulations \cite{MBG97} and kinetic theory calculations \cite{TSS98,TS94} show that it actually has a local minimum at that point.

Obviously, the Couette and Poiseuille flows can be combined to become the Couette--Poiseuille (or Poiseuille--Couette) flow \cite{CLL06,P66,ST06b,TK00}. To the best of our knowledge, all the studies on the Couette--Poiseuille flow assume that the Poiseuille part is driven by a  pressure gradient, not by an external force. This paper intends to fill this gap by considering the steady state of a dilute gas enclosed between two infinite parallel plates in relative motion, the particles of the gas being subject to the action of a uniform body force. This Couette--Poiseuille flow is sketched in Fig.\ \ref{fig1}(c). We will study the problem by the tools of kinetic theory by solving the BGK model for Maxwell molecules.

The aim of this work is two-fold. First, we want to investigate how the fully developed non-Newtonian Couette flow is distorted by the action of the external force. To that end we will assume a finite value of the Knudsen number related to the shear rate and perform a perturbation expansion to first order in the force. As a second objective, we will study how the non-Newtonian force-driven Poiseuille flow is modified by the shearing. This is done by assuming that the shear-rate Knudsen number and the scaled force are of the same order and neglecting terms of third and higher order. In both cases we are interested in the physical properties in the central {bulk} region of the slab, outside the influence of the boundary layers.

The organization of the paper is as follows. The Boltzmann equation for the Couette--Poiseuille flow is presented in Sec.\ \ref{sec2}. Section \ref{sec3} deals with the NS description of the problem. The main part of the paper is contained in Sec.\ \ref{sec3b}, where the kinetic theory approach is worked out. Some technical calculations are relegated to Appendix \ref{appA}. The results are graphically presented and discussed in Sec.\ \ref{sec6}. The paper ends with some concluding remarks in Sec.\ \ref{sec7}.

\section{The Couette--Poiseuille flow. Symmetry properties\label{sec2}}
Let us consider a dilute monatomic gas enclosed between two infinite parallel plates located at $y=\pm L/2$. The plates are in relative motion with velocities $U_\pm $ along the $x$ axis and are kept at a common temperature $T_w$. The imposed shear rate is therefore $\gd=(U_+-U_-)/L$. Besides, an external body force $\mathbf{F}=mg \widehat{\mathbf{x}}$, where $m$ is the mass of a particle and $g$ is a constant acceleration, is applied. The geometry of the problem is sketched in Fig.\ \ref{fig1}(c). In the absence of the external force ($g=0$) this problem reduces to the plane Couette flow [see Fig.\ \ref{fig1}(a)]. On the other hand, if the plates are at rest ($\gd=0$), one is dealing with the force-driven Poiseuille flow [see Fig.\ \ref{fig1}(b)]. The general problem with $\gd\neq 0$ and $g\neq 0$ defines the Couette--Poiseuille flow analyzed in this paper.

In the steady state only gradients along the $y$ axis are present and thus the Boltzmann equation becomes
\beq
\left(v_y\frac{\partial}{\partial y}+g\frac{\partial}{\partial v_x}\right) f(y,\mathbf{v}|\gd,g)=J[f,f],
\label{1}
\eeq
where $f$ is the one-particle velocity distribution function and $J[f,f]$ is the Boltzmann collision operator \cite{C88,C90,CC70,DvB77,K10,S05}, whose explicit expression will not be written down here. The notation $f(y,\mathbf{v}|\gd,g)$ emphasizes the fact that, apart from its spatial and velocity dependencies, the distribution function depends on the independent external parameters $\gd$ and $g$.  As said above, $g=0$ and $\gd=0$ correspond to the Couette and Poiseuille flows, respectively.

 In general, Eq.\ \eqref{1}  must be solved
subjected to specific  boundary conditions, which can be expressed in terms of the
kernels $K_{\pm}({\bf v}, {\bf v}')$ defined as follows. When a particle
with velocity ${\bf v}'$ hits the wall at $y=L/2$, the probability of being
reemitted with a velocity ${\bf v}$ within the range $d{\bf v}$ is
$K_{+}({\bf v}, {\bf v}')d{\bf v}$; the kernel $K_{-}({\bf v}, {\bf v}')$
represents the same but at $y=-L/2$.
The boundary conditions are then \cite{DvB77}
\beq
\Theta(\pm v_y)|v_y| f(y=\mp L/2,{\bf v})=\Theta(\pm v_y)\int d{\bf v}'\,
|v_y'|K_{\mp}({\bf v}, {\bf v}')\Theta(\mp v_y')f(y=\mp L/2,{\bf v}'),
\label{6.3:3.1}
\eeq
where $\Theta(x)$ is Heaviside's step function.
In the case of boundary conditions of complete accommodation with
the walls, so that $K_{\pm}({\bf v}, {\bf v}')=K_{\pm}({\bf v})$ does not
depend on the incoming velocity ${\bf v}'$, the kernels can be written as
\begin{equation}
K_{\mp}({\bf v})=A_{\mp}^{-1}\Theta(\pm v_y)|v_y| \varphi_{\mp}({\bf v}),\quad
A_{\mp}\equiv\int d{\bf v}\,\Theta(\pm v_y)|v_y| \varphi_{\mp}({\bf v}),
\label{6.3:3.2}
\end{equation}
where $\varphi_{\mp}({\bf v})$ represents the probability distribution of a
fictitious gas in contact with the system at $y=\mp L/2$. Equation
(\ref{6.3:3.2}) can then be interpreted as meaning that when a particle hits a
wall, it is instantaneously absorbed and  replaced by a particle leaving the fictitious
bath. Of course, any choice of $\varphi_{\mp}({\bf v})$ must be consistent with
the imposed wall velocities and temperatures, i.e.\
\begin{equation}
U_{\mp}=\int d{\bf v}\, v_x
\varphi_{\mp}({\bf v}),
\label{6.3:3.3}
\end{equation}
\begin{equation}
k_BT_{w}=\frac{1}{3}m\int d{\bf v}\, ({\bf v}-{\bf
U}_{\mp})^2
\varphi_{\mp}({\bf v}).
\label{6.3:3.4}
\end{equation}
Inserting Eq.\ (\ref{6.3:3.2}) into Eq.\ (\ref{6.3:3.1}), one has
\begin{equation}
\Theta(\pm v_y)f(y=\mp L/2,{\bf v})=\Theta(\pm v_y) n_{\mp} \varphi_{\mp}(\mathbf{v}),
\label{6.3:new1}
\end{equation}
where
\begin{equation}
n_{\mp}\equiv\frac{\int d\mathbf{v}'\, \Theta(\mp v_y')|v_y'| f(y=\mp L/2,\mathbf{v}')}{\int d\mathbf{v}'\, \Theta(\pm v_y')|v_y'| \varphi_\mp (\mathbf{v}')}.
\label{6.3:new2}
\end{equation}
The boundary conditions (\ref{6.3:new1}) are usually referred to as  \textit{diffuse} (or stochastic) boundary conditions.
The simplest and most common choice is that of a Maxwell--Boltzmann (MB)
distribution \cite{TTKB98}:
\begin{equation}
\varphi_{\mp}^{\text{MB}}({\bf v})=\left(\frac{m}{2\pi k_BT_{w}}\right)^{3/2}
\exp\left[-\frac{m({\bf v}-{\bf U}_{\mp})^2}{2k_BT_{w}}\right].
\label{6.3:3.5}
\end{equation}

The first few moments of $f$ define the densities of conserved quantities (mass, momentum, and temperature) and the associated fluxes. More explicitly,
\beq
n(y|\gd,g)=\int d\mathbf{v}\, f(y,\mathbf{v}|\gd,g),
\label{2}
\eeq
\beq
n(y|\gd,g)\mathbf{u}(y|\gd,g)=\int d\mathbf{v}\, \mathbf{v}f(y,\mathbf{v}|\gd,g),
\label{3}
\eeq
\beq
n(y|\gd,g)k_BT(y|g)=p(y|\gd,g)=\frac{m}{3}\int d\mathbf{v}\, V^2(y,\mathbf{v}|\gd,g)f(y,\mathbf{v}|\gd,g),
\label{4}
\eeq
\beq
P_{ij}(y|\gd,g)=m\int d\mathbf{v}\, V_i(y,\mathbf{v}|\gd,g)V_j(y,\mathbf{v}|\gd,g)f(y,\mathbf{v}|\gd,g),
\label{5}
\eeq
\beq
\mathbf{q}(y|\gd,g)=\frac{m}{2}\int d\mathbf{v}\, V^2(y,\mathbf{v}|\gd,g)\mathbf{V}(y,\mathbf{v}|\gd,g)f(y,\mathbf{v}|\gd,g).
\label{6}
\eeq
In these equations $n$ is the number density, $\mathbf{u}$ is the flow velocity,
\beq
\mathbf{V}(y,\mathbf{v}|a,\gd,)\equiv \mathbf{v}-\mathbf{u}(y|\gd,g)
\label{6b}
\eeq
is the peculiar velocity, $T$ is the temperature, $k_B$ is the Boltzmann constant, $p$ is the hydrostatic pressure, $P_{ij}$ is the pressure tensor, and $\mathbf{q}$ is the heat flux.
Taking velocity moments in both sides of Eq.\ \eqref{1} one gets the following \emph{exact} balance equations
\beq
\partial_y P_{yy}=0,
\label{7}
\eeq
\beq
\partial_y P_{xy}=mng,
\label{8}
\eeq
\beq
\partial_y q_y+P_{xy}\partial_y u_x=0.
\label{9}
\eeq
Henceforth, without loss of generality,  we will assume $u_x(0)=0$. In other words, we will adopt a reference frame solidary with the flow at the midpoint $y=0$.

The symmetry properties of the Couette--Poiseuille flow imply the following invariance properties of the velocity distribution function:
\beqa
\label{2.1}
f(y,v_x,v_y,v_z|\gd,g)&=&f(-y,-v_x,-v_y,v_z|\gd,-g) \nonumber\\
&=&f(-y,v_x,-v_y,v_z| -\gd,g)\nn
&=&f(y,v_x,v_y,-v_z|\gd,g),
\eeqa
As a consequence, if $\chi(y|\gd,g)$ denotes a hydrodynamic variable or a flux, one has
\beqa
\chi(y|\gd,g)&=&S_g \chi(-y|\gd,-g)\nn
&=& S_{\gd} \chi(-y|-\gd,g),
\label{10}
\eeqa
where $S_g=\pm 1$ and $S_{\gd}=\pm 1$. The parity factors $S_g$ and $S_{\gd}$ for the non-zero hydrodynamic fields and fluxes are displayed in Table \ref{table1}. In general, if $\chi$ is a moment of the form
\beq
\chi(y|\gd,g)=\int d\mathbf{v}\, V_x^{k_x}(y|\gd,g)v_y^{k_y}v_z^{2k_z} f(y,\mathbf{v}|\gd,g)
\label{11}
\eeq
then $S_g=(-1)^{k_x+k_y}$ and $S_{\gd}=(-1)^{k_y}$.

\begin{table}
\begin{tabular}{ccc}
\hline
Quantity& $S_g$&$S_{\gd}$\\
\hline
$n$&$+$&$+$\\
$u_x$&$-$&$+$\\
$T$&$+$&$+$\\
$p$&$+$&$+$\\
$P_{xx}$&$+$&$+$\\
$P_{yy}$&$+$&$+$\\
$P_{xy}$&$+$&$-$\\
$q_x$&$-$&$+$\\
$q_y$&$-$&$-$\\
\hline
\end{tabular}
\caption{Parity factors $S_g$ and $S_{\gd}$ for the hydrodynamic fields and the fluxes [see Eq.\ \protect\eqref{10}].}
\label{table1}
\end{table}

The general solution to the stationary Boltzmann equation (\ref{1}) with the  boundary conditions (\ref{6.3:new1}) can be split into two parts \cite{S91,S00,S02}:
\begin{equation}
f(y,\mathbf{v}|\gd,g)=f_H(y,\mathbf{v}|\gd,g)+f_B(y,\mathbf{v}|\gd,g).
\label{6.3:new3}
\end{equation}
Here, $f_H$ represents the \textit{hydrodynamic}, Hilbert-class, or  \textit{normal} contribution to the distribution function. This means that $f_H$ depends on space only through a \emph{functional} dependence on the  hydrodynamic fields, i.e.\
\begin{equation}
f_H(y)=f_H[n,T,u_x].
\label{6.3:new4}
\end{equation}
The contribution $f_B$ represents the  boundary-layer correction  to $f_H$, so that $f=f_H+f_B$ verifies the specified  boundary conditions. The correction $f_B$ is appreciably different from zero only in a thin layer (the so-called boundary layer  or Knudsen layer), adjacent to the plates, of thickness of the order of  the mean free path.
Consequently, if the separation $L$ between the plates is much larger than the characteristic mean free path, there exists a well defined  bulk region where the boundary correction vanishes and the distribution function is fully given by its hydrodynamic part.
In the boundary layers the hydrodynamic profiles are much less smooth than in the bulk domain.
The values of the flow velocity near the walls are different from the velocity of the plates (velocity slip phenomenon), i.e.\ $u_x(y=\pm L/2)\neq U_{\pm}$.  Besides,
the extrapolation of the velocity profile in the bulk to the boundaries, $u_{x,H}(y=\pm L/2)$, is also different from both the actual  values $u_x(y=\pm L/2)$ and the wall velocities $U_\pm$. Of course, an analogous  temperature jump effect takes place with the temperature profile.
The boundary contribution $f_B$ for small  Knudsen numbers has been analyzed elsewhere \cite{C90,S02}.

In the remaining of this paper we will focus on the hydrodynamic part $f_H$ (and will drop the subscript H), with special emphasis on the corresponding hydrodynamic contributions to the momentum and heat fluxes.
In order to nondimensionalize the problem, we choose quantities evaluated at the central plane $y=0$ as units:
\beq
f^*(s,\mathbf{v}^*|a,g^*)\equiv\frac{v_T^3(0)}{n(0)}f(y,\mathbf{v}|\gd,g),\quad \mathbf{v}^*\equiv\frac{\mathbf{v}}{v_T(0)},\quad  v_T(0)\equiv \sqrt{\frac{k_BT(0)}{m}},
\label{19}
\eeq
\beq
n^*(s|a,g^*)\equiv\frac{n(y|\gd,g)}{n(0)},\quad T^*(s|a,g^*)\equiv\frac{T(y|\gd,g)}{T(0)}, \quad p^*(s|a,g^*)\equiv\frac{p(y|\gd,g)}{p(0)},
\label{20}
\eeq
\beq
P_{ij}^*(s|a,g^*)\equiv\frac{P_{ij}(y|\gd,g)}{p(0)},\quad \mathbf{q}^*(s|a,g^*)\equiv\frac{\mathbf{q}(y|\gd,g)}{p(0)v_T(0)},
\eeq
\beq
a\equiv\frac{1}{\nu(0)}\left.\frac{\partial u_x}{\partial y}\right|_{y=0}, \quad g^*\equiv\frac{g}{v_T(0)\nu(0)},\quad \nu^*\equiv\frac{\nu}{\nu(0)} .
\label{20b}
\eeq
In the above equations we have found it convenient to introduce the dimensionless \emph{scaled} spatial variable
\beq
s(y)\equiv\frac{1}{v_T(0)}\int_0^y dy'\, \nu(y'),
\label{21}
\eeq
where $\nu(y)$ is an effective collision frequency. For the sake of concreteness, we choose it as
\beq
\nu(y)= \frac{p(y)}{\eta(y)},
\label{22}
\eeq
where $\eta$ is the NS shear viscosity.
The change from the boundary-imposed shear rate $\gd$ to the \emph{reduced local shear rate} $a$ is motivated by our goal of focusing on the central \emph{bulk} region of the system, outside the boundary layers.
Note that $a$ represents the \emph{Knudsen number} associated with the velocity gradient at $y=0$. Likewise, $g^*$  measures the strength of the external field on a particle moving
with the thermal velocity along a distance on the order of the mean free path.

The relationship \eqref{21} can be inverted to yield
\beq
y^*(s)=\int_0^s \frac{ds'}{\nu^*(s')},\quad y^*\equiv \frac{y}{v_T(0)/\nu(0)}.
\label{21b}
\eeq

The invariance properties \eqref{2.1} translate into
\beqa
\label{2.1b}
f^*(s,v_x^*,v_y^*,v_z^*|a,g^*)&=&f^*(-s,-v_x^*,-v_y^*,v_z^*|a,-g^*) \nonumber\\
&=&f^*(-s,v_x^*,-v_y^*,v_z^*|-a,g^*)\nn
&=&f^*(-s,v_x^*,v_y^*,-v_z^*|a,g^*).
\eeqa
Given the symmetry properties \eqref {2.1b}, we can restrict ourselves to $a>0$ and $g^*>0$ without loss of generality.

\section{Navier--Stokes description\label{sec3}}
To gain some insight into the type of fields one can expect in the Couette--Poiseuille flow, it is instructive to analyze the solution provided by the NS level of description. In the geometry of the problem, the NS constitutive equations are
\beq
P_{xx}=P_{yy}=P_{zz}=p,
\label{12}
\eeq
\beq
P_{xy}=-\eta\partial_y u_x,
\label{13}
\eeq
\beq
q_x=0,
\label{14}
\eeq
\beq
q_y=-\kappa \partial_y T,
\label{15}
\eeq
where $\eta$ is the shear viscosity, as said above, and $\kappa$ is the thermal conductivity. Inserting the NS \emph{approximate} relations \eqref{12}--\eqref{15} into the \emph{exact} conservation equations \eqref{7}--\eqref{9} one gets
\beq
p=\text{const},
\label{16}
\eeq
\beq
\left(\eta\partial_y\right)^2 u_x=-\eta mn g,
\label{17}
\eeq
\beq
\frac{5k_B}{2m\Pr}\left(\eta\partial_y\right)^2 T=-\left(\eta\partial_y u_x\right)^2,
\label{18}
\eeq
where $\Pr=(5k_B/2m)\eta/\kappa\simeq\frac{2}{3}$ is the Prandtl number. In dimensionless form, Eqs.\ \eqref{17} and \eqref{18} can be rewritten as
\beq
\partial_s^2 u_x^*(s)=-\frac{n^*(s)}{\nu^*(s)}g^*,
\label{23}
\eeq
\beq
\partial_s^2 T^*(s)=-\frac{2\Pr}{5}\left[\partial_s u_x^*(s)\right]^2.
\label{24}
\eeq
For simplicity, let us assume that the particles are Maxwell molecules \cite{C88,CC70,TM80}, so $\nu(y)\propto n(y)$ and $\nu^*(s)=n^*(s)$. In that case, Eqs.\ \eqref{23} and \eqref{24} allow for an explicit solution:
\beq
u_x^*(s|a,g^*)=a s-\frac{1}{2}g^*s^2,
\label{25}
\eeq
\beq
T^*(s|a,g^*)=1-\frac{\Pr}{30} s^2\left(6a^2-4 a g^* s+{g^*}^2 s^2\right).
\label{26}
\eeq
Here we have applied the Galilean choice $u_x(0)=0$ and the symmetry property $\left.\partial_y T\right|_{y=0}=0$.

Equation \eqref{25} shows that, according to the NS approximation, the velocity field in the Couette--Poiseuille flow is simply the superposition of the (quasi) linear Couette profile and the (quasi) parabolic Poiseuille profile. In the case of the temperature field, however, apart from the (quasi) parabolic Couette profile and the (quasi) quartic Poiseuille profile, a (quasi) cubic coupling term is present. Here we use the term ``quasi'' because the simple polynomial forms in Eqs.\ \eqref{25} and \eqref{26} refer to the scaled variable $s$. To go back to the real spatial coordinate $y$ one needs to make use of the relationship \eqref{21}, taking into account that for Maxwell molecules $\nu\propto n$. Instead of expressing $s$ as a function of $y$ it is  more convenient to proceed in the opposite sense by using Eq.\ \eqref{21b}. Since $1/\nu^*= T^*$ one simply has
\beq
y^*(s)=s\left[1-\frac{\Pr}{30} s^2\left(2a^2-a g^* s+\frac{1}{5}{g^*}^2 s^2\right)\right].
\label{27}
\eeq
For further use, note that, according to Eq.\ \eqref{26},
\beq
\left.\frac{\partial^2 T^*}{\partial {y^*}^2}\right|_{y^*=0}=-\Pr \frac{2}{5}a^2.
\label{NScurv}
\eeq
Thus, the NS temperature profile presents a \emph{maximum} at the midpoint $y^*=0$.

Before closing this section, let us write the pressure tensor and the heat flux profiles provided by the NS description:
\beq
P_{xx}^*(s|a,g^*)=P_{yy}^*(s|a,g)=P_{zz}^*(s|a,g)=1,
\label{NS1}
\eeq
\beq
P_{xy}^*(s|a,g^*)=-a+g^*s,
\label{NS2}
\eeq
\beq
q_{x}^*(s|a,g^*)=0,
\label{NS3}
\eeq
\beq
q_{y}^*(s|a,g^*)=s\left(a^2-ag^* s+\frac{1}{3}{g^*}^2 s^2\right).
\label{NS4}
\eeq

\section{Kinetic theory description. Perturbation solution.\label{sec3b}}

Now we want to get the hydrodynamic and flux profiles in the bulk domain of the system  from a purely kinetic approach, i.e., without assuming \emph{a priori} the applicability of the NS constitutive equations. To that end, instead of considering the detailed Boltzmann operator $J[f,f]$ we will make use of the celebrated BGK kinetic model \cite{BGK54,C88,W54}. In the BGK model Eq.\ \eqref{1} is replaced by
\beq
\left(v_y\frac{\partial}{\partial y}+g\frac{\partial}{\partial v_x}\right) f(y,\mathbf{v}|\gd,g)=-\nu(y|\gd,g)\left[f(y,\mathbf{v}|\gd,g)-\MM(y,\mathbf{v}|\gd,g)\right],
\label{28}
\eeq
where $\nu$ is the effective collision frequency defined by Eq.\ \eqref{22} and
\beq
\MM(\mathbf{v})=n\left(\frac{m}{2\pi k_BT}\right)^{3/2} \exp\left(-\frac{m V^2}{2k_BT}\right)
\label{29}
\eeq
is the local equilibrium distribution function. In terms of the dimensionless variables introduced in Eqs.\ \eqref{19}--\eqref{21}, Eq.\ \eqref{28} can be rewritten as
\beq
\left(1+v_y^*\partial_s\right)f^*(s,\mathbf{v}^*|a,g^*)=\MM^*(s,\mathbf{v}^*|a,g^*)-\frac{g^*}{\nu^*(s|a,g^*)}\frac{\partial}{\partial v_x^*}f^*(s,\mathbf{v}^*|a,g^*).
\label{30}
\eeq
Its \emph{formal} solution is
\beqa
f^*(\mathbf{v}^*)&=&\left(1+v_y^*\partial_s\right)^{-1}\left[\MM^*(\mathbf{v}^*)-\frac{g^*}{\nu^*}\frac{\partial}{\partial v_x^*}f^*(\mathbf{v}^*)\right]\nn
&=&\sum_{k=0}^\infty (-v_y^*\partial_s)^k \left[\MM^*(\mathbf{v}^*)-\frac{g^*}{\nu^*}\frac{\partial}{\partial v_x^*}f^*(\mathbf{v}^*)\right].
\label{35}
\eeqa
The formal character of the solution \eqref{35} is due to the fact that $f^*$ appears on the right-hand side  explicitly and also implicitly through $\MM^*$ and $\nu^*$. The solvability (or consistency) conditions are
\beq
\int d\mathbf{v}^*\left\{1,\mathbf{v}^*,{v^*}^2\right\}f^*(s,\mathbf{v}^*|a,g^*)=\int d\mathbf{v}^*\left\{1,\mathbf{v}^*,{v^*}^2\right\}\MM^*(s,\mathbf{v}^*|a,g^*).
\label{36}
\eeq

Let us assume now that $g^*$ is a small parameter so the solution to Eq.\ \eqref{30} can be expanded as
\beq
f^*(s,\mathbf{v}^*|a,g^*)=f_0^*(s,\mathbf{v}^*|a)+f_1^*(s,\mathbf{v}^*|a)g^*+f_2^*(s,\mathbf{v}^*|a){g^*}^2+\cdots.
\label{31}
\eeq
Likewise,
\beq
\chi^*(s|a,g^*)=\chi_0^*(s|a)+\chi_1^*(s|a)g^*+\chi_2^*(s|a){g^*}^2+\cdots,
\label{32}
\eeq
where $\chi^*$ denotes a generic velocity moment of $f^*$. The expansions of $n^*$, $\mathbf{u}^*$, and $T^*$ induce the corresponding expansion of $\MM^*$. The expansion in powers of $g^*$ allows the iterative solution of Eq.\ \eqref{35} by  a scheme similar to that followed in Ref.\ \cite{TGS99} in the case of an external force normal to the plates.

\subsection{Zeroth order in $g$. Pure Couette flow\label{sec4}}
\subsubsection{Finite shear rates}

To zeroth order in $g^*$, Eqs.\ \eqref{30} and \eqref{35} become
\beq
\left(1+v_y^*\partial_s\right)f_0^*(s,\mathbf{v}^*|a)=\MM_0^*(s,\mathbf{v}^*|a),
\label{38.0}
\eeq
\beq
f_0^*(s,\mathbf{v}^*|a)=\sum_{k=0}^\infty (-v_y^*\partial_s)^k \MM_0^*(s,\mathbf{v}^*|a),
\label{37}
\eeq
where
\beq
\MM_0^*(\mathbf{v}^*)=\frac{p_0^*}{\left(2\pi \right)^{3/2} {T_0^*}^{5/2}}\exp\left(-\frac{{V_0^*}^2}{2T_0^*}\right),
\quad \mathbf{V}_0^*\equiv \mathbf{v}^*-\mathbf{u}_0^*.
\label{33}
\eeq
These  are just the equations corresponding to the pure Couette flow. The complete solution has been obtained elsewhere \cite{BSD87,GS03,KDSB89} and so here we only quote the final results. The hydrodynamic profiles are
\beq
p_0^*(s|a)=1,
\label{38}
\eeq
\beq
u_{x,0}^*(s|a)= as,
\label{39}
\eeq
\beq
T_0^*(s|a)=1-\gamma(a) s^2,
\label{40}
\eeq
where the dimensionless parameter $\gamma(a)$ is a
{\em nonlinear\/} function of the reduced shear rate $a$ given implicitly
through the equation \cite{BSD87,GS03}
\begin{equation}
\label{2.10}
a^2=\gamma\left[3+2\frac{F_2(\gamma)}{F_1(\gamma)}\right],
\end{equation}
where the mathematical functions $F_r(x)$ are defined by
\beq
\label{2.12G}
F_0(x)=\frac{2}{x}\int_{0}^{\infty}dt\, t e^{-t^2/2}
K_0(2x^{-1/4}t^{1/2}), \quad F_r(x)=\left(\frac{d}{dx}x\right)^rF_0(x),
\eeq
$K_0(x)$ being the zeroth-order modified Bessel function.
Equation \eqref{2.12G} clearly shows that $F_r(x)$ has an essential singularity at $x=0$ and thus its  expansion in powers of $x$,
\beq
\label{2.11F}
F_r(x)=\sum_{k=0}^{\infty}
(k+1)^r(2k+1)!(2k+1)!!(-x)^k ,
\eeq
is asymptotic and not convergent. However, the series representation \eqref{2.11F} is Borel summable \cite{BSD87,KDSB89}, the corresponding integral representation  being given by Eq.\ \eqref{2.12G}.
The functions $F_r(x)$ with $r\geq 3$ can be easily
expressed in terms of $F_0(x)$, $F_1(x)$, and $F_2(x)$ as
\beq
\label{A14.1}
F_3(x)=\frac{1-F_0(x)}{8x}-F_2(x)-\frac{1}{4}F_1(x),
\eeq
\beq
\label{A14.2}
F_r(x)=\frac{1}{8x}\sum_{m=0}^{r-3}\binom{r-3}{m}(-1)^{m+r}F_m(x)-F_{r-1}(x)-\frac
{1}{4}F_{r-2}(x),
\quad r\geq 4.
\eeq

It is interesting to compare the hydrodynamic profiles with the results obtained from the Boltzmann equation at NS order (see Sec.\ \ref{sec3}). We observe that Eq.\ \eqref{38} agrees with Eq.\ \eqref{16} and Eq.\ \eqref{39} agrees with Eq.\ \eqref{25} for $g^*=0$. On the other hand, Eq.\ \eqref{26} with $g^*=0$ differs from Eq.\ \eqref{40}, except in the limit of small shear rates, in which case $\gamma(a)\approx \frac{1}{5}a^2$ (Note that $\Pr=1$ in the BGK model).

The relevant transport coefficients of the steady Couette flow are obtained
from the pressure tensor and the heat flux. They are highly nonlinear
functions of the reduced shear rate $a$ given by \cite{BSD87,GH97,GS03,MG98}
\begin{equation}
\label{2.13}
P_{xx,0}^*(s|a)= 1+4\gamma[F_1(\gamma)+F_2(\gamma)],
\end{equation}
\begin{equation}
\label{2.14}
P_{yy,0}^*(s|a)= 1-2\gamma[F_1(\gamma)+2F_2(\gamma)],
\end{equation}
\begin{equation}
\label{2.15}
P_{zz,0}^*(s|a)= 1-2\gamma F_1(\gamma),
\end{equation}
\begin{equation}
\label{2.17}
P_{xy,0}^*(s|a)= -a F_0(\gamma) ,
\end{equation}
\beq
\label{2.19}
q_{x,0}^*(s|a)=\frac{a}{2}\left\{F_0(\gamma)-1 -10\gamma F_1(\gamma)-8\gamma F_2(\gamma)\left[1-\frac{F_2(\gamma)}{F_1(\gamma)}\right]\right\}s,
\end{equation}
\begin{equation}
\label{2.18}
q_{y,0}^*(s|a)=a^2F_0(\gamma)s.
\end{equation}

Notice that, although the temperature gradient is only directed
along the $y$ axis (so that there is a response in this direction
through $q_y^*$), the shear flow induces a nonzero $x$ component of the heat
flux \cite{GH97,GS03,MG98,RC97}. Furthermore, normal stress differences (absent at NS order) are present. Equations \eqref{2.17} and \eqref{2.18} can be used to identify generalized nonlinear shear viscosity and thermal conductivity coefficients.

In general, the velocity moments of degree $k$ of $f^*_0$ are polynomial functions of the spatial variable $s$ of degree $k-2$. An explicit expression for the velocity distribution function $f^*_0$ has also been derived \cite{GS03,KDSB89}.

\subsubsection{Limit of small shear rates}
The coefficient $\gamma(a)$ characterizing the profile of the zeroth-order temperature $T_0^*$ is a complicated nonlinear function of the reduced shear rate $a$, as clearly apparent from Eq.\ \eqref{2.10}. Obviously, the zeroth-order pressure tensor and  heat flux  given by Eqs.\ \eqref{2.13}--\eqref{2.18} inherit this nonlinear character.

It is illustrative to assume that the reduced shear rate $a$ is small so one can express the quantities of interest as the first few terms in a (Chapman--Enskog) series expansion. {}From Eqs.\ \eqref{2.10}--\eqref{2.18} one obtains
\beq
\gamma(a)=\frac{a^2}{5}\left(1+\frac{72}{25}a^2+\cdots\right),
\label{56}
\eeq
\beq
P_{xx,0}^*(s|a)=1+\frac{8a^2}{5}\left(1-\frac{198}{25}a^2+\cdots\right),
\label{57}
\eeq
\beq
P_{yy,0}^*(s|a)=1-\frac{6a^2}{5}\left(1-\frac{228}{25}a^2+\cdots\right),
\label{58}
\eeq
\beq
P_{zz,0}^*(s|a)=1-\frac{2a^2}{5}\left(1-\frac{108}{25}a^2+\cdots\right),
\label{59}
\eeq
\beq
P_{xy,0}^*(s|a)=-a\left(1-\frac{18}{5}a^2+\cdots\right),
\label{60}
\eeq
\beq
q_{x,0}^*(s|a)=-\frac{14a^3}{5}\left(1-\frac{1836}{175}a^2+\cdots\right)s,
\label{61}
\eeq
\beq
q_{y,0}^*(s|a)=a^2\left(1-\frac{18}{5}a^2+\cdots\right)s.
\label{62}
\eeq

The terms of order $a^2$, $a$, and $a^2$ in Eqs. \eqref{56}, \eqref{60}, and \eqref{62}, respectively, agree with the corresponding NS expressions, Eqs.\ \eqref{26}, \eqref{NS2}, and \eqref{NS4}. On the other hand, as noted above, the normal stress differences ($P_{xx}^*-P_{yy}^*$ and $P_{zz}^*-P_{yy}^*$) and the streamwise heat flux component $q_x^*$ reveal non-Newtonian effects of orders $a^2$ and $a^3$, respectively.

\subsection{First order in $g$. Couette--Poiseuille flow\label{sec5}}
\subsubsection{Finite shear rates}
To first order in $g^*$ Eq.\ \eqref{35} yields
\beq
f_1^*(\mathbf{v}^*)-\MM_1^*(\mathbf{v}^*)=\Lambda^\aone(\mathbf{v}^*)+\Lambda^\bone(\mathbf{v}^*),
\label{41}
\eeq
where
\beq
\Lambda^\aone(\mathbf{v}^*)\equiv \sum_{k=1}^\infty (-v_y^*\partial_s)^k \MM_1^*(\mathbf{v}^*),\quad
\Lambda^\bone(\mathbf{v}^*)\equiv -\sum_{k=0}^\infty (-v_y^*\partial_s)^k T_0^*\frac{\partial}{\partial v_x^*}f_0^*(\mathbf{v}^*),
\label{41b}
\eeq
\beq
\MM_1^*(\mathbf{v}^*)=\MM_0^*(\mathbf{v}^*)\left[{p_1^*}+\frac{T_1^*}{2T_0^*}\left(\frac{{V_0^*}^2}{T_0^*}-5\right)+\frac{{u}_{x,1}^*
}{T_0^*}V_{x,0}^*\right],
\label{34}
\eeq
and we have already specialized to Maxwell molecules, so that $\nu^*=p^*/T^*$.
In order to apply the consistency conditions \eqref{36} in the derivation of the hydrodynamic fields $p_1^*$, $u_{x,1}^*$, and $T_1^*$,  it is convenient to define the moments
\beq
\Phi_{n_1n_2n_3}=\Phi_{n_1n_2n_3}^\aone+\Phi_{n_1n_2n_3}^\bone,
\label{46c}
\eeq
\beq
\Phi_{n_1n_2n_3}^\aone =\int d\mathbf{v}^*\, {V_{x,0}^*}^{n_1}{v_{y}^*}^{n_2}{v_{z}^*}^{n_3}\Lambda^\aone(\mathbf{v}^*),
\label{46a}
\eeq
\beq
\Phi_{n_1n_2n_3}^\bone =\int d\mathbf{v}^*\, {V_{x,0}^*}^{n_1}{v_{y}^*}^{n_2}{v_{z}^*}^{n_3}\Lambda^\bone(\mathbf{v}^*).
\label{46b}
\eeq
Therefore, the consistency conditions are
\beq
\Phi_{000}=0,
\label{42}
\eeq
\beq
\Phi_{100}=0,
\label{43}
\eeq
\beq
\Phi_{010}=0,
\label{44}
\eeq
\beq
\Phi_{200}+\Phi_{020}+\Phi_{002}=0.
\label{45}
\eeq

The evaluation of $\Phi_{n_1n_2n_3}^\aone$ and $\Phi_{n_1n_2n_3}^\bone$ is carried out in Appendix \ref{appA}.  The first-order profiles are
\beq
p_1^*(s|a)=\poo(a)s,
\label{A1}
\eeq
\beq
u_{x,1}^*(s|a)=\uot(a) s^2,
\label{A2}
\eeq
\beq
T_1^*(s|a)=\Tot(a) s^3,
\label{A3}
\eeq
where
\beq
\poo(a)=-\frac{1}{\gamma}\left(1-\frac{F_1}{F_0}\right)\Tot(a),
\label{46}
\eeq
\beq
\uot(a)=\frac{2\gamma\left(F_1+2F_2\right)-3}{6F_1}-\frac{a}{\gamma}\left(\frac{F_1}{F_0}-\frac{F_2}{F_1}\right)\Tot(a),
\label{47}
\eeq
\beq
\Tot(a)=\frac{4}{3}a\gamma^2 F_0\frac{4\gamma F_2\left(F_1+2F_2\right)-F_1\left(1-F_0\right)-6F_2}{D(a)},
\label{48}
\eeq
with
\beqa
D(a)&\equiv& 2\gamma F_1\left[F_0^2-2F_0+F_1-2\gamma\left(F_0-2F_1\right)\left(F_1+2F_2\right)\right]+a^2F_0F_1\left(1-F_0\right)\nn
&&-2a^2\gamma \left[F_0\left(F_1^2+6F_1F_2+8F_2^2\right)-2F_1^2\left(F_1+2F_2\right)\right].
\label{49}
\eeqa
In the above equations the functions $F_r$ are understood to be evaluated at $x=\gamma$.

As shown in Appendix \ref{appA}, the moment $\Phi_{n_1n_2n_3}$ is a polynomial function of $s$ of degree $n_1+n_2+n_3-1$. In particular, the non-zero elements of the first-order pressure tensor are
\beq
P_{ij,1}^*(s|a)=P_{ij,1}^{(1)}(a)s,
\label{63}
\eeq
with
\beq
P_{xx,1}^{(1)}(a)=3\poo(a)-P_{zz,1}^{(1)}(a),
\label{50}
\eeq
\beq
P_{yy,1}^{(1)}(a)=0,
\label{51}
\eeq
\beq
P_{zz,1}^{(1)}(a)=-\left(\frac{F_0-1}{2\gamma}+F_1+2F_2\right)\Tot-\left[2\gamma\left(F_1+2F_2\right)-1\right]\poo,
\label{52}
\eeq
\beq
P_{xy,1}^{(1)}(a)=1.
\label{53}
\eeq
As for the heat flux vector, the results are
\beq
q_{i,1}^*(a|s)=q_{i,1}^{(0)}(a)+q_{i,1}^{(2)}(a) s^2,
\label{54}
\eeq
where
\beqa
\label{3.1.9}
\qxoz(a)&=& a\left[\frac{2F_1-F_0-1}{8\gamma^2}+\frac{F_1+26F_2}{4\gamma}+a^2\left(\frac{1-F_0}{8\gamma^3}+\frac{7F_1-16F_2}{4\gamma^2}\right)\right]
 \Tot\nonumber \\
&&+\left(\frac{1-F_0}{4\gamma}+3F_2-\frac{1}{2}F_1+3a^2\frac{F_1-F_2}{\gamma}\right)\uot+a\left(\frac{1-F_0}{4\gamma}+3F_2\right.\nn
&&\left.-\frac{1}{2}F_1+a^2\frac{F_1-F_2}{\gamma}\right)\poo+\frac{2}{3}F_0+\frac{1}{6}F_1+\frac{5}{3}-
\frac{10\gamma}{3}(F_1+2F_2)\nn
&&+a^2\left(\frac{1-F_0}{4\gamma}+\frac{3}{2}F_1+3F_2\right),
\eeqa
\beqa
\label{3.1.8}
\qxot(a)&=&a\left[\frac{2F_0-F_2-1}{2\gamma}+F_1+2F_2-a^2\left(\frac{9F_0-2F_1-7}{8\gamma^2}\right.\right.\nn
&&\left.\left.+3\frac{F_1+14F_2}{4\gamma}\right)\right] \Tot-\left[\frac{3}{2}-\frac{1}{2}F_1-F_0+2\gamma(3F_1+4F_2)\right.\nn
&&\left. -3a^2\left(\frac{1-F_0}{4\gamma}-\frac{1}{2}F_1-3F_2\right)\right]\uot+a\left[-1+\frac{1}{2}F_0+\frac{1}{2}F_1\right.\nn
&&\left. +2\gamma(F_1+2F_2)+a^2\left(\frac{1-F_0}{4\gamma}+\frac{1}{2}F_1-5F_2\right)\right]\poo\nn
&&-\frac{\gamma}{6}\left(4+4F_0+5F_1+2F_2\right)+\frac{4}{3}\gamma^2(F_1+2F_2)+\frac{a^2}{2}\left(F_1-F_0\right),\nn
\eeqa
\beqa
\label{3.1.10}
\qyoz(a)&=& \left(\frac{1-F_0}{8\gamma^2}-\frac{7F_1+2F_2}{4\gamma}+a^2\frac{3F_1-F_2-2F_0}{2\gamma^2}\right)
 \Tot+a\frac{F_1-F_0}{\gamma}\uot\nonumber \\
&&-\left(\frac{3}{2}F_1+F_2
+a^2\frac{F_0-F_1}{2\gamma}\right)\poo-\frac{2a}{3}(2F_1+F_2),
\eeqa
\beqa
\label{3.1.11}
\qyot(a)&=& \left[\frac{F_1-1}{4\gamma}+\frac{1}{2}F_1+F_2+\frac{a^2}{4}\left(\frac{1-F_0}{2\gamma^2}-\frac{2F_0-7F_1+14F_2}{\gamma}\right)\right]
 \Tot\nn
 &&+a\left(F_0+F_1-2F_2\right)\uot
+\left[\frac{1}{4}(1-F_0)+\gamma(F_1+2F_2)\right.\nn
&&\left.
-\frac{a^2}{2}(F_0-3F_1+2F_2)\right]\poo+\frac{a}{6}\left[1-F_0-2\gamma(F_1+2F_2)\right]
.
\eeqa

Equations \eqref{51} and \eqref{53} are consistent with the momentum balance equations \eqref{7} and \eqref{8}, respectively. The energy balance equation \eqref{9} requires that
\beq
\qyot=a\left(F_0\uot-\frac{1}{2}\right).
\label{55}
\eeq
Taking into account Eqs.\ \eqref{46}--\eqref{48}, it is possible to check that Eqs.\ \eqref{3.1.11} and \eqref{55} are indeed equivalent.

Let us now get the relationship between the scaled space variable $s$ and the true (dimensionless) coordinate $y^*$. {}From the definition \eqref{21} we have
\beqa
\frac{dy^*}{ds}&=&\frac{T^*(s|a,g^*)}{p^*(s|a,g^*)}\nn
&=&T_0^*(s|a)+\left[T_1^*(s|a)-T_0^*(s|a)p_1^*(s|a)\right]g^*+\mathcal{O}({g^*}^2).
\label{73}
\eeqa
Inserting Eqs.\ \eqref{40}, \eqref{A1}, and \eqref{A3} one gets
\beq
y^*(s|a,g^*)=s-\frac{\gamma(a)}{3}s^3-\frac{s^2}{4}\left\{2\poo(a)-\left[\Tot(a)+\gamma(a)\poo(a)\right]s^2\right\}g^*+\mathcal{O}({g^*}^2).
\label{74}
\eeq

\subsubsection{Limit of small shear rates}
As done in the case of $g^*=0$, it is illustrative to obtain the first-order coefficients \eqref{46}--\eqref{48}, \eqref{50}, \eqref{52}, and \eqref{3.1.9}--\eqref{3.1.11} in the limit of small shear rates. Taking into account Eqs.\ \eqref{2.11F} and \eqref{56} one gets
\beq
\poo(a)=-\frac{12a}{5}\left(1-\frac{73}{5}a^2+\cdots\right),
\label{64}
\eeq
\beq
\uot(a)=-\frac{1}{2}-\frac{29a^2}{5}\left(1-\frac{23136}{725}a^2+\cdots\right),
\label{65}
\eeq
\beq
\Tot(a)=\frac{2a}{15}\left(1+\frac{109}{5}a^2+\cdots\right),
\label{66}
\eeq
\beq
P_{xx,1}^{(1)}(a)=-\frac{28a}{5}\left(1-\frac{431}{35}a^2+\cdots\right),
\label{67}
\eeq
\beq
P_{zz,1}^{(1)}(a)=-\frac{8a}{5}\left(1-\frac{113}{5}a^2+\cdots\right),
\label{68}
\eeq
\beq
\qxoz(a)=-1+\frac{216a^2}{5}\left(1-\frac{23329}{225}a^2+\cdots\right),
\label{69}
\eeq
\beq
\qxot(a)=\frac{29a^2}{5}\left(1-\frac{1844}{145}a^2+\cdots\right),
\label{70}
\eeq
\beq
\qyoz(a)=-\frac{19a}{5}\left(1-\frac{8172}{95}a^2+\cdots\right),
\label{71}
\eeq
\beq
\qyot(a)=-a\left(1+4a^2+\cdots\right).
\label{72}
\eeq

According to Eqs.\ \eqref{25}, \eqref{26}, and \eqref{NS4}, the NS equations yield $\uot=-\frac{1}{2}$, $\Tot=\frac{2}{15}a$ (taking $\Pr=1$), and $\qyot=-a$, so that they agree with the leading terms in Eqs.\ \eqref{65}, \eqref{66}, and \eqref{72}. On the other hand, the NS approximation fails in accounting for the non-zero values of $\poo$, $P_{xx,1}^{(1)}$, $P_{zz,1}^{(1)}$, $\qxoz$, $\qxot$, and $\qyoz$. In particular, $\qxoz\neq 0$ even in the pure Poiseuille flow ($a=0$) \cite{S09,ST06,TSS98,TS94}.

\section{Results and discussion\label{sec6}}
\subsection{Finite shear rates. First order in $g^*$}

\begin{figure}[tbp]
  \includegraphics[width=\columnwidth]{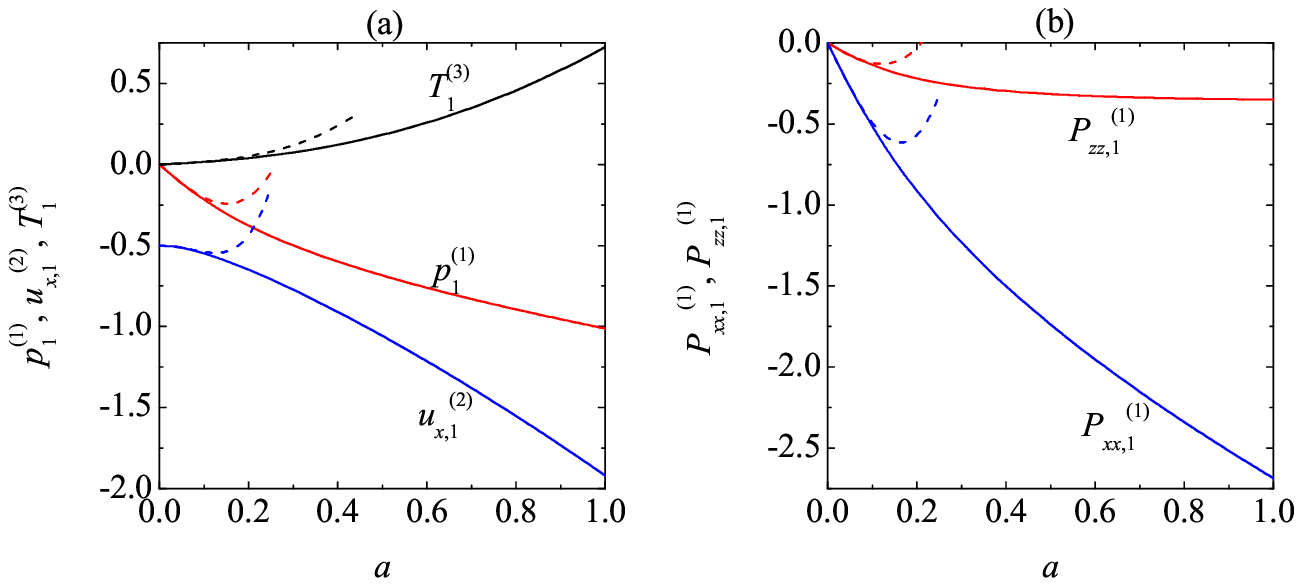}
\begin{center}
\caption{First-order coefficients $\poo$, $\uot$, $\Tot$ (left panel), $P_{xx,1}^{(1)}$, and $P_{zz,1}^{(1)}$ (right panel) as functions of the reduced shear rate $a$. The dashed lines represent the terms shown in Eqs.\ \protect\eqref{64}--\protect\eqref{68}.}\label{fig2}
  \end{center}
\end{figure}

\begin{figure}[tbp]
  \includegraphics[width=\columnwidth]{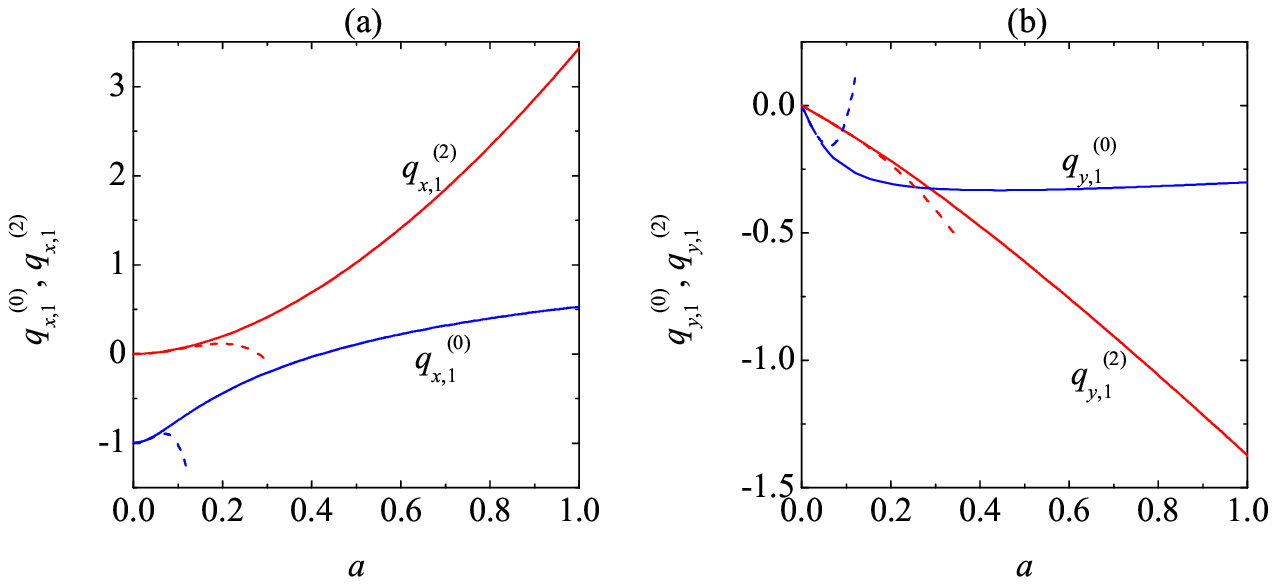}
\begin{center}
\caption{First-order coefficients $\qxoz$, $\qxot$ (left panel), $\qyoz$, and $\qyot$ (right panel) as functions of the reduced shear rate $a$. The dotted lines represent the terms shown in Eqs.\ \protect\eqref{69}--\protect\eqref{72}.}\label{fig3}
  \end{center}
\end{figure}

The nonlinear dependence on the reduced shear rate $a$ of the zeroth-order quantities has been analyzed elsewhere \cite{BSD87,GH97,GS03}, so that here we focus on the first-order corrections.
Figure \ref{fig2}(a) shows the coefficients associated with the hydrodynamic profiles, i.e., $\poo(a)$, $\uot(a)$, and $\Tot(a)$. The first two quantities are negative, while the third one is positive, in agreement with what might be expected in view of Eqs.\ \eqref{64}--\protect\eqref{66}. On the other hand,  the practical range of applicability of the truncated series \eqref{64}--\protect\eqref{66} is restricted to small shear rates ($a\lesssim 0.1$). The addition of further terms in the (Chapman--Enskog) expansion in powers of $a$ would not improve that range because of the asymptotic character of the series. Note that the range of applicability of the NS description (according to which $\poo=0$, $\uot=-\frac{1}{2}$, and $\Tot=\frac{2}{15}a$) is even much more restrictive, especially in the case of $\poo$.

The normal-stress coefficients $P_{xx,1}^{(1)}(a)$ and $P_{zz,1}^{(1)}(a)$ are plotted in Fig.\ \ref{fig2}(b). Both coefficients vanish in the NS description. Again, the truncated series \eqref{67} and \eqref{68} are reliable only for $a\lesssim 0.1$. We observe that the $xx$ element has a much larger magnitude than the $zz$ element. The other relevant coefficients of the pressure tensor are not plotted  because they are identically  $P_{yy,1}^{(1)}=0$ and $P_{yy,1}^{(1)}=1$, as a consequence of the exact momentum balance equations \eqref{7} and \eqref{8}.

The coefficients associated with the heat flux vector are plotted in Fig.\ \ref{fig3}. According to the NS equations, $\qxoz=\qxot=\qyoz=0$ and $\qyot=-a$, what strongly contrasts with the nonlinear behavior observed in Fig.\ \ref{fig3}. This is especially dramatic in the case of the streamwise coefficient $\qxot$, which deviates from zero even in the limit $a\to 0$. It is interesting to note that, while $\qxot$, $\qyoz$, and $\qyot$ have definite signs (at least in the interval $0\leq a\leq 1$), the coefficient $\qxoz$ changes from negative to positive around $a\simeq 0.42$.

\begin{figure}[tbp]
  \includegraphics[width=\columnwidth]{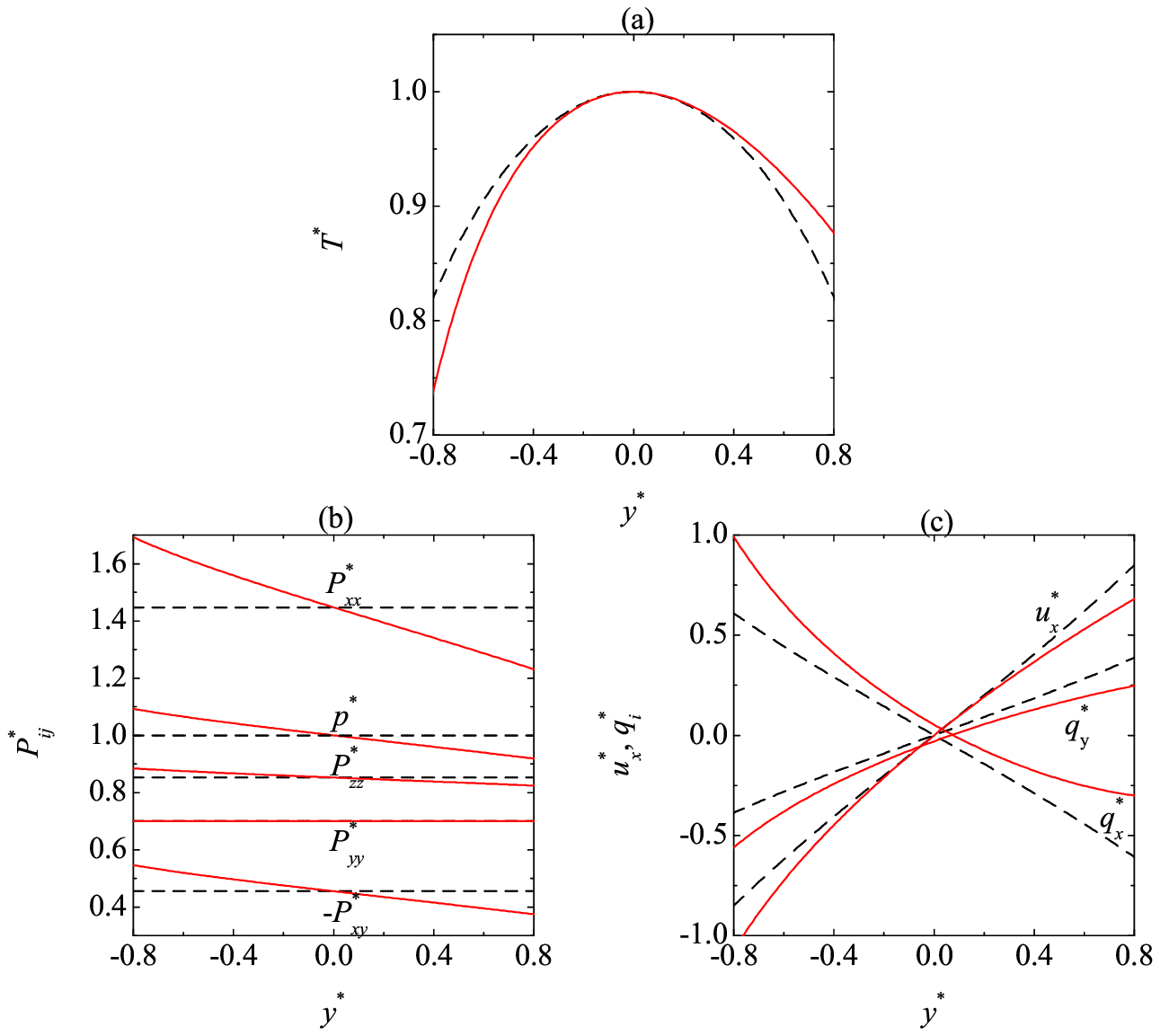}
\begin{center}
\caption{Profiles of (a) temperature, (b) elements of the pressure tensor, and (c) flow velocity and components of the heat flux vector. The value of the reduced shear rate is $a=1$. Two values of the external force are considered: $g^*=0$ (dashed lines) and $g^*=0.1$ (solid lines).}\label{fig4}
  \end{center}
\end{figure}

In order to illustrate how the Couette-flow profiles are distorted by the action of the external force, we will take $a=1$ with $g^*=0$ and  $g^*=0.1$.
While the latter value is possibly not small enough as to make the first-order calculations sufficient, our aim here is to highlight the trends
to be expected when the fully nonlinear Couette flow coexists with the force-driven Poiseuille flow. The profiles are shown in Fig.\ \ref{fig4}. In the pure Couette flow ($g^*=0$) the temperature and pressure profiles are symmetric, while the velocity and heat flux profiles are antisymmetric. The application of the external force breaks these symmetry features since the first-order terms have symmetry properties opposite to those of the zeroth-order terms, in agreement with the signs of $S_g$ in Table \ref{table1}. As a consequence, the temperature gradient increases across the channel with respect to that of the Couette flow, as shown by Fig.\ \ref{fig4}(a). The elements of the pressure tensor are no longer uniform but exhibit negative gradients, especially in the case of the normal stress $P_{xx}^*$ [see Fig.\ \ref{fig4}(b)]. An exception is $P_{yy}^*$, which is exactly uniform as a consequence of the momentum balance equation \eqref{7}. To first order in $g^*$ the value of  $P_{yy}^*$ is the same as in the Couette flow [see Eq.\ \eqref{51}], but this situation changes when terms of order ${g^*}^2$ are added \cite{S09,TSS98,TS94}. We observe from Fig.\ \ref{fig4}(c) that the flow velocity (in the reference frame moving with the midplane $y=0$) is decreased by the action of the external force. A similar behavior is presented by $q_y^*$, what qualitatively correlates with the increase in the temperature gradient observed in Fig.\ \ref{fig4}(a). Regarding the component  $q_x^*$, it takes larger values in the Couette--Poiseuille flow than in the pure Couette flow. In both cases ($g^*=0$ and $g^*=0.1$) the shearing is so large ($a=1$) that the two components of the heat flux have a similar magnitude, i.e., $|q_x^*|\sim |q_y^*|$. An interesting effect induced by the external field is the existence of a non-zero heat flux at $y^*=0$, even though the  temperature gradient vanishes at that point. More specifically, $q_x^*(0)=\qxoz>0$ and $q_y^*(0)=\qyoz<0$. The sign of the former quantity  changes, as noted before,  at $a\simeq 0.42$.

\subsection{Small shear rates. Second order in $g^*$}

Thus far, all the results are valid for arbitrary values of
the reduced shear rate $a$ but are restricted to first order in the reduced external field $g^*$. One could continue the perturbation scheme devised in Sec.\ \ref{sec3b} to further orders in $g^*$ but the analysis becomes extremely cumbersome if one still wants to keep $a$ arbitrary. Furthermore, the perturbation expansion in powers of $g^*$ is expected to be only asymptotic \cite{TS94}.
On the other hand, we can combine the results obtained here [see Eqs.\ \eqref{56}--\eqref{62} and \eqref{64}--\eqref{72}] with those of Ref.\ \cite{TS94} to obtain the hydrodynamic and flux profiles to second order in $g^*$, by assuming that both the reduced shear rate and the reduced force are of the same order, i.e., $a\sim g^*$. The results are
\beq
p^*(s|a,g^*)=1-\frac{12}{5}a g^* s+\frac{6}{5}{g^*}^2 s^2+\cdots,
\label{75}
\eeq
\beq
u_x^*(s|a,g^*)=as-\frac{1}{2} g^* s^2+\cdots,
\label{76}
\eeq
\beq
T^*(s|a,g^*)=1-\frac{1}{5}a^2s^2+\frac{2}{15}a g^* s^3+{g^*}^2s^2\left(\frac{19}{25}-\frac{1}{30}s^2\right)+\cdots,
\label{77}
\eeq
\beq
P_{xx}^*(s|a,g^*)=1+\frac{8}{5} a^2-\frac{28}{5}a g^* s+{g^*}^2\left(\frac{328}{25}+\frac{14}{5}s^2\right)+\cdots,
\label{78}
\eeq
\beq
P_{yy}^*(s|a,g^*)=1-\frac{6}{5} a^2-\frac{306}{25}{g^*}^2+\cdots,
\label{79}
\eeq
\beq
P_{zz}^*(s|a,g^*)=1-\frac{2}{5} a^2-\frac{8}{5}a g^* s-{g^*}^2\left(\frac{22}{25}-\frac{4}{5}s^2\right)+\cdots,
\label{80}
\eeq
\beq
P_{xy}^*(s|a,g^*)=-a+ g^* s+\cdots,
\label{81}
\eeq
\beq
q_{x}^*(s|a,g^*)=- g^*+\cdots,
\label{82}
\eeq
\beq
q_{y}^*(s|a,g^*)=a^2s-a g^*\left(\frac{19}{5}+s^2\right)+\frac{1}{3}{g^*}^2s^3+\cdots.
\label{83}
\eeq
In these equations the ellipses denote terms that are at least of orders $a^3$, ${g^*}^3$, $a^2 g^*$ or $a{g^*}^2$. As for the relationship between $y^*$ and $s$, from Eqs.\ \eqref{21} or \eqref{21b} we get
\beq
s={y^*}+\frac{1}{15}a^2 {y^*}^3-\frac{1}{30}a g^* {y^*}^2\left(36+{y^*}^2\right)+\frac{1}{150}{g^*}^2 {y^*}^3\left(22+{y^*}^2\right)+\cdots.
\label{84}
\eeq
Therefore, we can safely replace $s$ by $y^*$ in Eqs.\ \eqref{75}--\eqref{83}.

\begin{figure}[tbp]
  \includegraphics[width=\columnwidth]{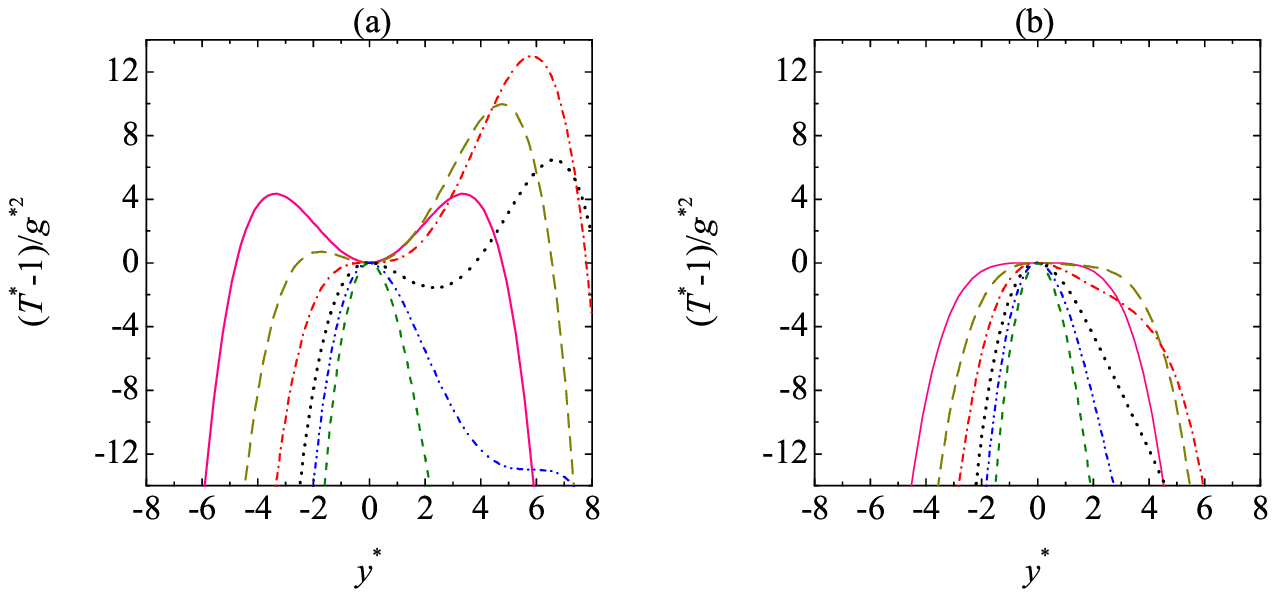}
\begin{center}
\caption{Profiles of the scaled temperature difference $(T^*-1)/{g^*}^2$  for $|g^*|\ll 1$ and several values of the ratio $a/g^*$: $a/g^*=0$ (---), $a/g^*=1$ (-- -- --), $a/g^*=\sqrt{19/5}$ (- $\cdot$ - $\cdot$ - $\cdot$), $a/g^*=3$ ($\cdots$), $a/g^*=2\sqrt{19/5}$ (- $\cdot\cdot$ - $\cdot\cdot$ - $\cdot\cdot$), and $a/g^*=5$ (- - - -). The left panel corresponds to the kinetic-theory predictions, Eq.\ \protect\eqref{85}, while the right panel corresponds to the NS predictions, Eq.\ \protect\eqref{26}.}\label{fig5}
  \end{center}
\end{figure}

Again, it is instructive to compare Eqs.\ \eqref{75}--\eqref{83} against the  NS predictions worked out in Sec.\ \ref{sec3} (with $\Pr=1$, in consistency with the BGK value of the Prandtl number). As can be seen from  Eqs.\ \eqref{16}, \eqref{25}, \eqref{26}, and \eqref{NS1}--\eqref{NS4}, the NS expressions do not contain terms of order higher than $a^2$, ${g^*}^2$, or $a{g^*}$ and therefore their comparison with the kinetic-theory results \eqref{75}--\eqref{83} is not biased.
We see that only the NS results for $u_x^*$ and $P_{xy}^*$ are supported by kinetic theory. This does not mean that Newton's law for the shear stress, Eq.\ \eqref{13} is satisfied, since $\eta=p/\nu$ and the hydrostatic pressure $p$ is not actually uniform.  As said before, the NS constitutive equations also fail to account for the existence of normal stress differences (typical non-Newtonian effects) as well as of a streamwise component of the heat flux (failure of Fourier's law).  Perhaps  the most interesting and subtle differences refer to the presence in Eqs.\ \eqref{77} and \eqref{83} of the extra terms $\frac{19}{25}{g^*}^2s^2$ and $-\frac{19}{5}a g^*$, respectively, which are absent in their NS counterparts, Eqs.\ \eqref{26} and \eqref{NS4}. The extra term in $q_y^*$ implies that $q_y^*(0)\neq 0$, what represents a clear violation of Fourier's law \eqref{15}. The term $\frac{19}{25}{g^*}^2s^2$ present in the temperature field \eqref{77} has dramatic consequences on the curvature of the temperature profile at the midpoint $y^*=0$. {}From Eq.\ \eqref{77} we get
\beq
\left.\frac{\partial^2 T^*}{\partial {y^*}^2}\right|_{y^*=0}=-\frac{2}{5}a^2+\frac{38}{25}{g^*}^2.
\label{curv}
\eeq
Therefore, while  the NS temperature profile presents a maximum at $y^*=0$ [see Eq.\ \eqref{NScurv}], Eq.\ \eqref{curv} shows that the profile has actually a local \emph{minimum} if $a^2<\frac{19}{5}{g^*}^2$. To analyze this feature in more detail, let us rewrite Eq.\ \eqref{77} as
\beq
\frac{T^*-1}{{g^*}^2}\approx-\frac{1}{5}\left(\frac{a}{g^*}\right)^2{y^*}^2+\frac{2}{15}\left(\frac{a}{g^*}\right) {y^*}^3+{y^*}^2\left(\frac{19}{25}-\frac{1}{30}{y^*}^2\right).
\label{85}
\eeq
Figure \ref{fig5}(a) shows the scaled temperature difference ($T^*-1)/{g^*}^2$, as given by Eq.\ \eqref{85}, for $a/g^*=0$, $1$, $\sqrt{19/5}$, $3$, $2\sqrt{19/5}$, and $5$. In the case $a/g^*=0$ (pure Poiseuille flow) the temperature profile has a minimum at $y^*=0$ surrounded by two symmetric maxima at $y^*=\pm 2\sqrt{19/5}$. When $a/g^*\neq 0$ (Couette--Poiseuille flow) several possibilities arise. If $0<a/g^*<\sqrt{19/5}$ one still has a local minimum at $y^*=0$ but the two maxima are no longer symmetric: the one with $y^*<0$ moves to the center, while the one with $y^*>0$ departs from it. This is represented by the case $a/g^*=1$ in Fig.\ \ref{fig5}(a). At  $a/g^*=\sqrt{19/5}$, the left maximum and the central minimum merge to become an inflection point of zero slope. Next, in the range $\sqrt{19/5}<a/g^*<2\sqrt{19/5}$ the temperature presents a local maximum at $y^*=0$ followed by a minimum and an absolute maximum, both with $y^*>0$. This situation is illustrated by the case $a/g^*=3$ in Fig.\ \ref{fig5}(a). At $a/g^*=2\sqrt{19/5}$ the minimum and maximum with $y^*>0$ merge to create an inflection point of zero slope. Finally, if $a/g^*>\sqrt{19/5}$ [see case $a/g^*=5$ in Fig.\ \ref{fig5}(a)] only the central maximum remains and the profile becomes more and more symmetric as $a/g^*$ increases. In the limit $a/g^*\to\infty$ (or, equivalently, $g^*\to 0$) one recovers the pure Couette flow. This rich phenomenology is absent in the case of the NS temperature profile, as shown in Fig.\ \ref{fig5}(b).

Given the physical interest of Eq.\ \eqref{77} or, equivalently, Eq.\ \eqref{85}, it is convenient to rewrite it in real units. This yields
\beqa
T(y)&=&T(0)-\frac{\eta(0)}{2\kappa(0)}\left(\frac{\partial u_x}{\partial y}\right)_{y=0}^2y^2+\frac{mn(0)}{3\kappa(0)}\left(\frac{\partial u_x}{\partial y}\right)_{y=0} g y^3\nn
&&-
\frac{m^2n^2(0)}{12\eta(0)\kappa(0)}g^2 y^4+C_T\frac{m^2}{k_B^2 T(0)}g^2 y^2,
\label{86}
\eeqa
where $C_T=\frac{19}{25}$ in the BGK model, while $C_T\simeq 1.0153$ in the Boltzmann equation for Maxwell molecules \cite{S09,ST06,TSS98}. Equation \eqref{86} still holds in the NS description, except that $C_T=0$.

\section{Concluding remarks\label{sec7}}
In this paper we have studied the stationary Couette--Poiseuille flow of a dilute gas. As illustrated by Fig.\ \ref{fig1}(c), the gas is enclosed between two infinite parallel plates in relative motion (Couette flow) and at the same time the particles feel the action of a uniform longitudinal force (force-driven Poiseuille flow) along the same direction as the moving plates. Our main goal has been to assess the limitations of the NS description of the problem and highlight the importance of non-Newtonian properties.

In order to get explicit results, the complicated Boltzmann collision operator has been replaced by the mathematically much simpler BGK model with a temperature-independent collision frequency (Maxwell molecules). The kinetic model has been solved to first order in the reduced force parameter $g^*$ for arbitrary values of the reduced shear rate $a$. Moreover, complementing these results with those obtained in previous works for the pure Poiseuille flow to second order in $g^*$, we have been able to get the solution to second order in both $a$ and $g^*$.

Starting from the pure nonlinear Couette flow, we have studied the influence of a weak external force, as measured by the nonlinear shear-rate dependence of the nine coefficients $\poo(a)$, $\uot(a)$, $\Tot(a)$, $P_{xx,1}^{(1)}(a)$, $P_{zz,1}^{(1)}(a)$, $\qxoz(a)$, $\qxot(a)$, $\qyoz(a)$, and $\qxot(a)$. These functions are plotted in Figs.\ \ref{fig2} and \ref{fig3}. A more intuitive picture on the distortion produced by the force on the Couette profiles for the hydrodynamic fields ($p$, $u_x$, and $T$), the pressure tensor ($P_{xx}$, $P_{yy}$, $P_{zz}$, and $P_{xy}$), and the heat flux ($q_x$ and $q_y$)  is provided by Fig.\ \ref{fig4}.

Complementarily,  we have obtained the quantities  of interest [cf.\ Eqs.\ \eqref{75}--\eqref{83}] when the shear rate and the force are treated on the same footing, both to second order. This has allowed us to analyze [see Fig.\ \ref{fig5}(a)] how, by starting from the pure Poiseuille flow, the symmetric bimodal temperature profile is strongly distorted by the shearing until arriving at the symmetric parabola characteristic of the pure Couette flow.

To put the paper in a proper context, it must be stressed that in our treatment we have focused on the  nonlinear bulk properties of the Couette--Poiseuille flow,   while ignoring linear effects like
Knudsen boundary layers. The latter effects, however, may be relevant in  applications like slow MEMS flows where the nonlinear effects considered
here might be less important.

Considering the great current interest in the force-driven Poiseuille flow as a playground to test hydrodynamic theories and theoretical approaches, we expect that the work presented here may contribute to motivate further studies, both theoretical and computational, on the Couette--Poiseuille flow.

\section*{Acknowledgments}
We dedicate this paper to the fond memory of Carlo Cercignani.
The work of A.S. has been supported by the
Ministerio de Ciencia e Innovaci\'on (Spain) through Grant No.\ FIS2010-16587 (partially financed by FEDER funds) and by the Junta
de Extremadura (Spain) through Grant No.\ GR10158.


\appendix

\section{Evaluation of $\Phi_{n_1n_2n_3}^\aone$ and $\Phi_{n_1n_2n_3}^\bone$\label{appA}}
{}From Eqs.\ \eqref{41b} and \eqref{46a} one gets
\beq
\label{2.9}
\Phi_{n_1n_2n_3}^\aone= \sum_{\ell=0}^{n_1}\sum_{k=\theta_\ell}^\infty\binom{k+\ell}{\ell}
 \frac{n_1!(-a)^\ell}{(n_1-\ell)!} ( - \partial_s)^k\int d\mathbf{v}^*
{V_{x,0}^*}^{n_1-\ell}{v_{y}^*}^{n_2+k+\ell}{v_{z}^*}^{n_3}
\MM_1^*(\mathbf{v}^*),
\eeq
where $\theta_0=1$ and $\theta_\ell=0$ for $\ell\geq 1$. In Eq.\ \eqref{2.9} use has been made of the mathematical relations
\beq
\label{2.10b}
A(s)( - \partial_s)^kB(s) =\sum_{\ell=0}^k \binom{k}{\ell}( - \partial_s)^{k-\ell}\left\{ \left[\partial_s^\ell A(s)\right]B(s)\right\},
\eeq
\beq
 \partial_s^\ell{V_{x,0}^*}^{n_1} =\frac{n_1!(-a)^\ell}{(n_1-\ell)!}{V_{x,0}^*}^{n_1-\ell} .
 \label{2.10c}
\eeq
Next, using the Maxwellian integrals
\beq
\label{2.11}
  \int d\mathbf{v}^*
{V_{x,0}^*}^{n_1}{v_{y}^*}^{n_2}{v_{z}^*}^{n_3}
\MM_{0}^*(\mathbf{v}^*)= K_{n_1}K_{n_2}K_{n_3}{T_0^*}^{(n_1+n_2+n_3-2)/2},
\eeq
where
$K_n=(n-1)!!$ if $n=\text{even}$  [with the convention  $(-1)!!=1$], being zero if $n=\text{odd}$,  Eq.\ (\ref{2.9}) becomes
\beqa
\label{2.11a}
\Phi_{n_1n_2n_3}^\aone&=& K_{n_3}\sum_{\ell=0}^{n_1}\sum_{k=\theta_\ell}^\infty\binom{k+\ell}{\ell}\frac{n_1!(-a)^\ell}{(n_1-\ell)!}K_{k+n_2+\ell} ( - \partial_s)^k
{T_0^*}^{(k+n_1+n_2+n_3-2)/2}\nn
&&\times\left\{K_{n_1-\ell}\left[p_1^*+\frac{k+n_1+n_2+n_3-2}{2}\frac{T_1^*}{T_0^*}\right]
+K_{n_1-\ell+1}\frac{u_{x,1}^*}{\sqrt{T_0^*}}\right\}.\nn
\eeqa

Before considering the integral $\Phi_{n_1,n_2,n_3}^\bone$, it is convenient to make use of the relation
\beq
(-\partial_s)^k \left[A(s)B(s)\right]=\sum_{m=0}^k\binom{k}{m}\left[(-\partial_s)^m A(s)\right]\left[(-\partial_s)^{k-m} B(s)\right]
\eeq
to rewrite the function $\Lambda^\bone(\mathbf{v}^*)$ as
\beq
\label{2.12}
\Lambda^\bone(\mathbf{v}^*)=(1-\gamma s^2)\Lambda^\bonezero(\mathbf{v}^*)+2\gamma s \Lambda^\boneone(\mathbf{v}^*)-2\gamma \Lambda^\bonetwo(\mathbf{v}^*),
\eeq
where
\beqa
\Lambda^\bonem(\mathbf{v}^*)&\equiv& -\frac{\partial}{\partial v_x^*}\sum_{k=0}^\infty \binom{k+m}{m}{v_y^*}^{k+m}(-\partial_s)^k f_0^*(\mathbf{v}^*)\nn
&=&-\frac{\partial}{\partial v_x^*}\sum_{k=0}^\infty \binom{k+m+1}{m+1}{v_y^*}^{k+m}(-\partial_s)^k \MM_0^*(\mathbf{v}^*).
\label{2.12a}
\eeqa
In the last step we have made use of Eq.\ \eqref{37} and the mathematical property
\beq
\label{2.14b}
\sum_{\ell=0}^k \binom{\ell+m}{m}=\binom{k+m+1}{m+1}.
\eeq

Insertion of Eq.\ \eqref{2.12} into Eq.\ \eqref{46b} gives
\beq
\label{2.12b}
\Phi_{n_1n_2n_3}^\bone=(1-\gamma s^2)\Phi_{n_1n_2n_3}^\bonezero+2\gamma s \Phi_{n_1n_2n_3}^\boneone-2\gamma \Phi_{n_1n_2n_3}^\bonetwo,
\eeq
where
\beq
\label{2.16.0}
\Phi_{n_1n_2n_3}^\bonem=\int d\mathbf{v}^*{V_{x,0}^*}^{n_1}{v_{y}^*}^{n_2}{v_{z}^*}^{n_3} \Lambda^\bonem(\mathbf{v}^*) .
\eeq
Using again Eqs.\ \eqref{2.10b}--\eqref{2.11}, one gets
\beqa
\label{2.16}
\Phi_{n_1n_2n_3}^\bonem&=& K_{n_3}\sum_{\ell=0}^{n_1-1}\sum_{k=0}^\infty\binom{k+\ell+m+1}{m+1}\binom{k+\ell}{\ell}\frac{n_1!(-a)^\ell}{(n_1-1-\ell)!}K_{n_1-1-\ell}\nn
&&\times K_{k+n_2+\ell+m} ( - \partial_s)^k
{T_0^*}^{(k+n_1+n_2+n_3+m-3)/2}.
\eeqa

Once the integrals $\Phi_{n_1n_2n_3}^\aone$ and $\Phi_{n_1n_2n_3}^\bone$ are expressed in terms of $T_0^*(s)$, $p_1^*(s)$, $u_{x,1}^*(s)$, and $T_1^*(s)$, we can apply the consistency conditions \eqref{42}--\eqref{45} to get the hydrodynamic profiles to first order in $g^*$. To that end, we first guess the polynomial forms \eqref{A1}--\eqref{A3},
so that only the coefficients $\poo$, $\uot$, and $\Tot$ remain to be determined. It is straightforward to check that $\Phi_{000}^\aone=\Phi_{000}^\bone=0$, and thus Eq.\ \eqref{42} is identically satisfied. The remaining relevant quantities in Eqs.\ \eqref{43}--\eqref{45} turn out to be
\beq
\Phi_{100}^\aone=2a\Tot\frac{F_1-F_2}{\gamma}+2(\uot+a \poo)F_1,
\label{2.19.0}
\eeq
\beq
\Phi_{010}^\aone=\Tot\frac{F_1-F_0}{\gamma}- \poo F_0,
\label{A4}
\eeq
\beqa
s^{-1}\Phi_{200}^\aone&=&\Tot\left[\left(\frac{10F_2-F_1}{\gamma}+\frac{F_0-1}{2\gamma^2}\right)a^2-
F_1-2F_2-\frac{F_0-1}{2\gamma}\right]\nn
&&+2\poo\left[\left(2F_2-F_1\right)a^2-\gamma\left(2F_2+F_1\right)\right]+8\uot a F_2,
\label{A6}
\eeqa
\beq
s^{-1}\Phi_{020}^\aone=\Tot\frac{F_0-F_1}{\gamma}-\poo\left(1-F_0\right),
\label{A7}
\eeq
\beq
s^{-1}\Phi_{002}^\aone=-\Tot\left(\frac{F_0-1}{2\gamma}+F_1+2F_2\right)-2\poo\gamma\left(F_1+2F_2\right),
\label{A8}
\eeq
\beq
\label{2.23}
\Phi_{100}^\bonezero=\frac{1}{T_0^*},\quad \Phi_{100}^\boneone=0,\quad \Phi_{100}^\bonetwo=\frac{1}{3}\left(F_1+2F_2\right),
\eeq
\beq
\label{A5}
\Phi_{010}^\bonezero=\Phi_{010}^\boneone=\Phi_{010}^\bonetwo=0,
\eeq
\beq
\label{A9}
\Phi_{200}^\bonezero=0,\quad \Phi_{200}^\boneone=-2a \left(F_1+2F_2\right),
\eeq
\beq
\label{A10}
\Phi_{200}^\bonetwo=-2a s\left(F_1+2F_2+\frac{F_0-1}{6\gamma}\right),
\eeq
\beq
\label{A11}
\Phi_{020}^\bonezero=\Phi_{020}^\boneone=\Phi_{020}^\bonetwo=0,
\eeq
\beq
\label{A12}
\Phi_{002}^\bonezero=\Phi_{002}^\boneone=\Phi_{002}^\bonetwo=0.
\eeq
In the above equations use has been made of Eqs.\ \eqref{2.11F}--\eqref{A14.2}.

Inserting Eqs.\ \eqref{A4} and \eqref{A5} into the consistency condition \eqref{44} one simply gets Eq.\ \eqref{46}. Next, insertion of Eqs.\ \eqref{2.19.0} and \eqref{2.23} into Eq.\ \eqref{43} allows one to obtain Eq.\ \eqref{47}. Finally, use of Eqs.\ \eqref{A6}--\eqref{A8} and \eqref{A9}--\eqref{A12} in Eq.\ \eqref{45} yields Eq.\ \eqref{48}.
Note that from Eqs.\ \eqref{46} and \eqref{A7} one gets $s^{-1}\Phi_{020}^\aone=-\poo$.

Taking into account that $T_0^*$, $p_1^*$, $u_{x,1}^*$, and $T_1^*$ are polynomials in $s$ of degrees $2$, $1$, $2$, and $3$, respectively, Eqs.\ \eqref{2.11a} and \eqref{2.16} show that $\Phi_{n_1n_2n_3}^\aone$ and $\Phi_{n_1n_2n_3}^\bonem$ are polynomials of degrees $n_1+n_2+n_3-1$ and $n_1+n_2+n_3+m-3$, respectively. Consequently, the moments defined by Eq.\ \eqref{46c} are polynomials of degree $n_1+n_2+n_3-1$.

Let us proceed now to the evaluation of the integrals $\Phi_{n_1n_2n_3}^\aone$ and  $\Phi_{n_1n_2n_3}^\bone$ related to the pressure tensor and the heat flux. The integrals related to the diagonal elements of the pressure tensor are already given by Eqs.\ \eqref{A6}--\eqref{A8} and \eqref{A9}--\eqref{A12}. For instance, $P_{yy,1}^*=p_1^*+\Phi_{020}$, with  similar relations for $P_{xx,1}^*$ and $P_{zz,1}^*$. The results are displayed in Eqs.\ \eqref{63}--\eqref{52}. In the case of the shear stress, one has $P_{xy,1}^*=\Phi_{110}$. {}From Eqs.\ \eqref{2.11a} and \eqref{2.16} we obtain
\beq
s^{-1}\Phi_{110}^\aone=\Tot a\frac{F_0-3F_1+2F_2}{\gamma}-2\uot F_1-\poo a(2F_1-F_0),
\label{A13}
\eeq
\beq
\Phi_{110}^\bonezero=0,\quad \Phi_{110}^\bonezero=F_1,\quad \Phi_{110}^\bonetwo=\frac{2}{3}(F_1-F_2)s.
\label{A14}
\eeq
This gives
\beqa
P_{xy,1}^*(a|s)&=&\left[a\frac{F_0-3F_1+2F_2}{\gamma}\Tot-2F_1\uot- a(2F_1-F_0)\poo\right.\nn
&&\left.+\frac{2}{3}\gamma(F_1+2F_2)\right]s.
\label{A17}
\eeqa
Making use of Eqs.\ \eqref{46} and \eqref{47}, it is easy to check that Eq.\ \eqref{A17} reduces to Eq.\ \eqref{53}.

In the case of the heat flux, one has
\beq
q_{x,1}^*=\frac{1}{2}\left(\Phi_{300}+\Phi_{120}+\Phi_{102}\right)-\left(P_{xx,0}^*-1\right)u_{x,1}^*,
\label{A15}
\eeq
\beq
q_{y,1}^*=\frac{1}{2}\left(\Phi_{210}+\Phi_{030}+\Phi_{012}\right)-P_{xy,0}^*u_{x,1}^*.
\label{A16}
\eeq
After tedious algebra one obtains Eqs.\ \eqref{54}--\eqref{3.1.11}.

\medskip
Received September 2010; revised October 2010.
\medskip


\begin{thebibliography}{99}

\bibitem{AC10}
\newblock M. Alam and V. K. Chikkadi,
\newblock \emph{Velocity distribution function and correlations in a granular Poiseuille flow},
\newblock J. Fluid Mech., \textbf{653} (2010), 175--219.

\bibitem{AS92}
\newblock M. Alaoui and A. Santos,
\newblock \emph{Poiseuille flow driven by an external force},
\newblock Phys. Fluids A, \textbf{4} (1992), 1273--1282.

\bibitem{ATN02}
\newblock K. Aoki, S. Takata and T. Nakanishi,
\newblock \emph{A Poiseuille-type flow of a rarefied gas between two parallel plates driven by a uniform external force},
\newblock Phys. Rev. E, \textbf{65} (2002), 026315.

\bibitem{AMN79}
\newblock E. Asmolov, N. K. Makashev and V. I. Nosik,
\newblock  \emph{Heat transfer between parallel plates in a gas of Maxwellian molecules},
\newblock  Sov. Phys. Dokl., \textbf{24} (1979), 892--894.

\bibitem{BGK54}
\newblock P. L. Bhatnagar, E. P. Gross and M. Krook,
\newblock \emph{A model collision processes in gases. I. Small amplitude processes in charged and neutral one-component systems},
\newblock {Phy. Rev.}, \textbf{94} (1954), 511--525.


\bibitem{BSD87}
\newblock J. J. Brey, A. Santos and J. W. Dufty,
\newblock \emph{Heat and momentum transport far from equilibrium},
\newblock Phys. Rev. A, \textbf{36}  (1987), 2842--2849.



\bibitem{C88}
\newblock C. Cercignani,
\newblock ``The Boltzmann Equation and Its Applications,''
\newblock Springer--Verlag, New York, 1988.


\bibitem{C90}
\newblock C. Cercignani,
\newblock ``Mathematical Methods in Kinetic Theory,''
\newblock Plenum Press, New York, 1990.

\bibitem{CLL06}
\newblock C. Cercignani, M. Lampis and S. Lorenzani,
\newblock \emph{Plane Poiseuille--Couette problem in micro-electro-mechanical systems applications with gas-rarefaction effects},
\newblock Phy. Fluids, \textbf{18} (2006), 087102.

\bibitem{CS66}
\newblock C. Cercignani and F. Sernagiotto,
\newblock \emph{Cylindrical Poiseuille flow of a rarefied gas},
\newblock Phys. Fluids, \textbf{9} (1966), 40--44.

\bibitem{CC70}
\newblock S. Chapman and T. G. Cowling,
\newblock {``The Mathematical Theory of Nonuniform Gases,''}
\newblock Cambridge University Press, Cambridge, 1970.

\bibitem{CA09}
\newblock V. Chikkadi and M. Alam,
\newblock \emph{Slip velocity and stresses in granular Poiseuille flow via event-driven simulation},
\newblock Phys. Rev. E, \textbf{80} (2009), 021303.

\bibitem{DvB77}
\newblock J. R. Dorfman and H. van Beijeren,
\newblock  \emph{The kinetic theory of gases},
\newblock  in ``Statistical Mechanics, Part B'' (ed.\ B. J. Berne),
                Plenum Press, New York, (1977), 65--179.




\bibitem{EF07}
\newblock  A. I. Erofeev and O. G. Friedlander,
\newblock \emph{Macroscopic models for non-equilibrium flows of monatomic gas and normal solutions},
\newblock in ``Rarefied Gas Dynamics: 25th International Symposium on Rarefied Gas Dynamics,''
 (eds. M. S. Ivanov and A. K. Rebrov),
 Publishing House of the Siberian Branch of the Russian Academy of Sciences, Novosibirsk, (2007),  117--124.

\bibitem{ELM94}
\newblock R. Esposito, J. L. Lebowitz and R. Marra,
\newblock \emph{A hydrodynamic limit of the stationary Boltzmann equation in a slab},
\newblock  Commun. Math. Phys., \textbf{160} (1994), 49--80.

\bibitem{GTRTS06}
\newblock M. A. Gallis, J. R. Torczynski, D. J. Rader, M. Tij and A. Santos,
\newblock \emph{Normal solutions of the Boltzmann equation for highly nonequilibrium Fourier flow and Couette flow},
\newblock Phys. Fluids, \textbf{18}  (2006), 017104.

\bibitem{GTRTS07}
\newblock M. A. Gallis, J. R. Torczynski, D. J. Rader, M. Tij and A. Santos,
\newblock \emph{Analytical and numerical normal solutions of the Boltzmann equation for highly nonequilibrium Fourier and Couette flows},
\newblock in ``Rarefied Gas Dynamics: 25th International Symposium on Rarefied Gas Dynamics,''
 (eds. M. S. Ivanov and A. K. Rebrov),
 Publishing House of the Siberian Branch of the Russian Academy of Sciences, Novosibirsk, (2007),  251--256.


\bibitem{GVU08}
\newblock L. S. Garc\'ia-Col\'in, R. M. Velasco and F. J. Uribe,
\newblock \emph{Beyond the Navier--Stokes equations: Burnett hydrodynamics},
\newblock Phys. Rep., \textbf{465} (2008), 149--189.

\bibitem{GH97}
\newblock V. Garz\'o and M. L\'opez de Haro,
\newblock \emph{Nonlinear transport for a dilute gas in steady Couette flow},
\newblock Phys. Fluids, {\bf 9} (1997), 776--787.


\bibitem{GS03}
\newblock V. Garz\'o and A. Santos,
\newblock ``Kinetic Theory of Gases in Shear Flows. Nonlinear Transport,''
\newblock Kluwer Academic Publishers, Dordrecht, 2003.

\bibitem{HM99}
\newblock S. Hess and M. Malek Mansour,
\newblock \emph{Temperature profile of a dilute gas undergoing a plane Poiseuille flow},
\newblock Physica A, \textbf{272} (1999), 481--496.

\bibitem{KMZ87}
\newblock   L. P. Kadanoff, G. R.  McNamara and G. Zanetti,
\newblock   \emph{A Poiseuille viscometer for lattice gas automata},
\newblock    Complex Syst., \textbf{1} (1987), 791--803.

\bibitem{KMZ89}
\newblock   L. P. Kadanoff, G. R.  McNamara and G. Zanetti,
\newblock   \emph{From automata to fluid flow: Comparisons of simulation and theory},
\newblock Phys. Rev. A, \textbf{40} (1989), 4527--4541.

\bibitem{KDSB89a}
\newblock   C. S. Kim, J. W. Dufty, A. Santos and J. J. Brey,
\newblock  \emph{Hilbert-class or ``normal'' solutions for stationary heat flow},
\newblock Phys. Rev. A, \textbf{39} (1989), 328--338.


\bibitem{KDSB89}
\newblock C. S. Kim, J. W. Dufty, A. Santos and J. J. Brey,
\newblock \emph{Analysis of nonlinear transport in Couette flow},
\newblock Phys. Rev A, \textbf{40} (1989), 7165--7174.

\bibitem{K10}
\newblock G. M. Kremer,
\newblock ``An Introduction to the Boltzmann Equation and Transport Processes in Gases,''
\newblock Springer, Berlin, 2010.

\bibitem{MBG97}
\newblock M. Malek Mansour, F. Baras and A. L. Garcia,
\newblock \emph{On the validity of hydrodynamics in plane Poiseuille flows},
\newblock Physica A, \textbf{240} (1997), 255--267.

\bibitem{MN81}
\newblock N. K. Makashev and V. I. Nosik,
\newblock \emph{Steady Couette flow (with heat transfer) of a gas of Maxwellian molecules},
\newblock  Sov. Phys. Dokl., \textbf{25} (1981), 589--591.


\bibitem{MASG94}
\newblock J. M. Montanero, M. Alaoui, A. Santos and V. Garz\'o,
\newblock \emph{Monte Carlo simulation of the Boltzmann equation for steady Fourier flow},
\newblock Phys. Rev. E, \textbf{49} (1994), 367--375.

\bibitem{MG98}
\newblock J. M. Montanero and V. Garz\'o,
\newblock \emph{Nonlinear Couette flow in a dilute gas: Comparison between theory and molecular dynamics simulation},
\newblock Phys. Rev. E, {\bf 58} (1998), 1836--1842.

\bibitem{MSG00}
\newblock J. M. Montanero, A. Santos and V. Garz\'o,
\newblock \emph{Monte Carlo simulation of nonlinear Couette flow in a dilute gas},
\newblock Phys. Fluids, \textbf{12} (2000), 3060--3073.

\bibitem{M09}
\newblock R. S. Myong,
\newblock \emph{Coupled nonlinear constitutive models for rarefied and microscale gas flows: subtle interplay of kinematics and dissipation effects},
\newblock Cont. Mech. Thermodyn., \textbf{21} (2009), 389--399.

\bibitem{N81}
\newblock V. I. Nosik,
\newblock \emph{Heat transfer between parallel plates in a mixture of gases of Maxwellian molecules},
\newblock  Sov. Phys. Dokl., \textbf{25} (1981), 495--497.

\bibitem{N83}
\newblock V. I. Nosik,
\newblock \emph{Degeneration of the Chapman--Enskog expansion in one-dimensional motions of Maxwellian molecule gases},
\newblock in ``Rarefied Gas Dynamics,'' vol.\ \textbf{13} (eds. O. M. Belotserkovskii, M. N. Kogan, S. S. Kutateladze,  and A. K.  Rebrov),
 Plenum Press, New York, (1983), 237--244.

\bibitem{OSA89}
\newblock T. Ohwada, Y. Sone and K. Aoki,
\newblock \emph{Numerical analysis of the Poiseuille and thermal transpiration flows between two parallel plates on the basis of the Boltzmann equation for hard-sphere molecules},
\newblock Phys. Fluids A, \textbf{1} (1989), 2042--2049.

\bibitem{P66}
\newblock M. C. Potter,
\newblock \emph{Stability of plane Couette--Poiseuille flow},
\newblock J. Fluid Mech.,  \textbf{24} (1966), 609--619.

\bibitem{RC97}
\newblock D. Risso and P. Cordero,
\newblock \emph{Dilute gas Couette flow: Theory and molecular dynamics simulation},
\newblock Phys. Rev. E, {\bf 56} (1997), 489--498.

\bibitem{RC98}
\newblock D. Risso and P. Cordero,
\newblock \emph{Generalized hydrodynamics for a Poiseuille flow: theory and simulations},
\newblock  Phys. Rev. E, \textbf{58} (1998), 546--553.

\bibitem{STS03}
\newblock M. Sabbane, M. Tij and A. Santos,
\newblock \emph{Maxwellian gas undergoing a stationary Poiseuille flow in a pipe},
\newblock Physica A, \textbf{327} (2003), 264--290.


\bibitem{S09}
\newblock A. Santos,
\newblock \emph{Solutions of the moment hierarchy in the kinetic theory of Maxwell models},
\newblock Cont. Mech. Thermodyn., \textbf{21} (2009), 361--387.


\bibitem{SBG86}
\newblock A. Santos, J. J. Brey and V. Garz\'o,
\newblock \emph{Kinetic model for steady heat flow},
\newblock Phys. Rev. A, \textbf{34} (1986), 5047--5050.

\bibitem{SBKD89}
\newblock A. Santos, J. J. Brey, C. S. Kim and J. W. Dufty,
\newblock \emph{Velocity distribution for a gas with steady heat flow},
\newblock Phys. Rev. A, \textbf{39} (1989), 320--327.

\bibitem{SGB92}
\newblock A. Santos, V. Garz\'o and J. J. Brey,
\newblock \emph{Comparison between the homogeneous-shear and the sliding-boundary methods to produce shear flow},
\newblock Phys. Rev. A, \textbf{46} (1992), 8018--8020.

\bibitem{ST06}
\newblock A. Santos and M. Tij,
\newblock \emph{Gravity-driven Poiseuille flow in dilute gases. Elastic and inelastic collisions},
\newblock  in ``Modelling and Numerics of Kinetic Dissipative Systems'' (eds.\ L. Pareschi, G. Russo and G. Toscani),
                Nova Science Publishers, New York, (2006), 53--67.

\bibitem{S91}
\newblock Y. Sone,
\newblock \emph{Asymptotic theory of a steady flow of a rarefied
				gas past bodies for small Knudsen numbers},
\newblock  in ``Advances in Kinetic Theory and Continuum
				Mechanics'' (eds.\ R. Gatignol and Soubbaramayer),
                Springer--Verlag, Berlin, (1991), 19--31.

\bibitem{S00}
\newblock Y. Sone,
\newblock \emph{Flows induced by temperature fields in a
				rarefied gas and their ghost effect on the
				behavior of a gas in the continuum limit},
\newblock Annu. Rev. Fluid Mech., \textbf{32} (2000), 779--811.

\bibitem{S02}
\newblock Y. Sone,
\newblock {``Kinetic Theory and Fluid Dynamics,''}
\newblock Birkh{\"{a}}user, Boston, 2002.


\bibitem{S05}
\newblock H. Struchtrup,
\newblock ``Macroscopic Transport Equations for Rarefied Gas Flows. Approximation Methods in Kinetic Theory,''
\newblock Springer, Berlin, 2005.

\bibitem{ST08}
\newblock H. Struchtrup and M Torrilhon,
\newblock \emph{Higher-order effects in rarefied channel flows},
\newblock  Phys. Rev. E, \textbf{78} (2008), 046301.

\bibitem{ST06b}
\newblock S. A. Suslov and T. D. Tran,
\newblock \emph{Revisiting plane Couette--Poiseuille flows of a piezo-viscous fluid},
\newblock J. Non-Newton. Fluid Mech., \textbf{154} (2006), 170--178.

\bibitem{TTS09}
\newblock P. Taheri, M. Torrilhon and H. Struchtrup,
\newblock  \emph{Couette and Poiseuille microflows: analytical solutions for regularized 13-moment equations},
\newblock  Phys. Fluids, \textbf{21} (2009), 017102.

\bibitem{TTKB98}
\newblock R. Tehver, F. Toigo,  J. Koplik and	J. R. Banavar,
\newblock \emph{Thermal walls in computer simulations},
\newblock Phys. Rev. E, \textbf{57} (1998), R17--R20.

\bibitem{TK00}
\newblock E. M. Thurlow and J. C. Klewicki,
\newblock \emph{Experimental study of turbulent Poiseuille--Couette flow},
\newblock Phys. Fluids, \textbf{12} (2000), 865--875.

\bibitem{TGS99}
\newblock M. Tij, V. Garz\'o and A. Santos,
\newblock \emph{Influence of gravity on nonlinear transport in the planar Couette flow},
\newblock Phys. Fluids, \textbf{11} (1999), 893--904.


\bibitem{TSS98}
\newblock M. Tij, M. Sabbane and A. Santos,
\newblock \emph{Nonlinear Poiseuille flow in a gas},
\newblock Phys. Fluids, \textbf{10} (1998), 1021--1027.




\bibitem{TS94}
\newblock M. Tij and A. Santos,
\newblock \emph{Perturbation analysis of a stationary nonequilibrium flow generated by an external force},
\newblock J. Stat. Phys., \textbf{76} (1994), 1399--1414.


\bibitem{TS95}
\newblock M. Tij and A. Santos,
\newblock \emph{Combined heat and momentum transport in a dilute gas},
\newblock Phys. Fluids, \textbf{7} (1995), 2858--2866.


\bibitem{TS01}
\newblock M. Tij and A. Santos,
\newblock \emph{Non-Newtonian Poiseuille flow of a gas in a pipe},
\newblock Physica A, \textbf{289} (2001), 336--358.


\bibitem{TS04}
\newblock M. Tij and A. Santos,
\newblock \emph{Poiseuille flow in a heated granular gas},
\newblock J. Stat. Phys., \textbf{117} (2004), 901--928.








\bibitem{TTMGSD01}
\newblock M. Tij, E. E. Tahiri, J. M. Montanero, V. Garz\'o, A. Santos and J. W. Dufty,
\newblock \emph{Nonlinear Couette flow in a low density granular gas},
\newblock J. Stat. Phys., \textbf{103} (2001), 1035--1068.

\bibitem{TE97}
\newblock B. D. Todd and D. J. Evans,
\newblock \emph{Temperature profile for Poiseuille flow},
\newblock Phys. Rev. E, \textbf{55} (1997), 2800--2807.

\bibitem{TTE97}
\newblock K. P. Travis,  B. D. Todd and D. J. Evans,
\newblock \emph{Poiseuille flow of molecular fluids},
\newblock Physica A, \textbf{240} (1997), 315--327.

\bibitem{TM80}
\newblock C. Truesdell and R. G. Muncaster,
\newblock ``Fundamentals of Maxwell's Kinetic Theory of a Simple Monatomic Gas,''
\newblock  Academic Press, New York, 1980.

\bibitem{UG99}
\newblock F. J. Uribe and A. L. Garcia,
\newblock \emph{Burnett description for plane Poiseuille flow},
\newblock Phys. Rev. E, \textbf{60} (1999), 4063--4078.

\bibitem{VSG10}
\newblock F. Vega Reyes, A. Santos and V. Garz\'o,
\newblock \emph{Non-Newtonian granular hydrodynamics. What do the inelastic simple shear flow and the elastic Fourier flow have in common?},
\newblock Phys. Rev. Lett., \textbf{104} (2010), 028001.


\bibitem{W54}
\newblock P. Welander,
\newblock \emph{On the temperature jump in a rarefied gas},
\newblock Akiv f\"or Fysik, \textbf{7} (1954), 507--553.

\bibitem{X03}
\newblock K. Xu,
\newblock \emph{Super-Burnett solutions for Poiseuille flow},
\newblock Phys. Fluids, \textbf{15} (2003), 2077--2080.

\bibitem{ZGA02}
\newblock Y. Zheng, A. L.  Garcia and B. J. Alder,
\newblock \emph{Comparison of kinetic theory and hydrodynamics for Poiseuille flow},
\newblock  J. Stat. Phys., \textbf{109} (2002), 495--505.


\end{thebibliography}
\end{document}